\documentclass[aps,pra, reprint,twocolumn]{revtex4-2}
\usepackage{amsmath}
\usepackage{amsfonts}
\usepackage{graphicx}
\usepackage{bm}
\usepackage[dvipsnames]{xcolor}
\usepackage{physics}

\newcommand{\bb}[1]{\mathbf{#1}}
\newcommand{\m}[1]{\mathcal{#1}}
\usepackage{ulem}
\usepackage{tikz-cd} 
\usepackage{tikz}

\usepackage{url}

\usepackage{mathtools}

\allowdisplaybreaks

\usepackage[colorlinks=true,linkcolor=blue,citecolor=red]{hyperref}

\begin{document}

\title {Dynamics Near a Photonic Band-Edge: Strong Coupling Effects Beyond Rotating-Wave Approximation}

\author{Matthieu Vanhoecke,\textsuperscript{1} Orazio Scarlatella,\textsuperscript{2} and Marco Schir\`o\textsuperscript{1}}
\affiliation{\textsuperscript{1} JEIP, UAR 3573 CNRS, Coll\`ege de France, PSL Research University, F-75321 Paris, France}
\affiliation{\textsuperscript{2} Clarendon Laboratory, University of Oxford, Parks Road, Oxford OX1 3PU, United Kingdom}
%

\date{\today}

\begin{abstract}
We study the dynamics of a quantum emitter coupled to a two-dimensional photonic crystal featuring a finite bandwidth with sharp edges and a Van-Hove singularity. We study the effect of strong system-bath coupling and non-Markovianity of the photonic environment using a non-perturbative approach based on the recently introduced NCA dynamical map for open quantum systems. We show that several characteristic features of the dynamics near a photonic band-edge such as the freezing of spontaneous emission and the maximum light-matter entanglement, get strongly modified in presence of counter-rotating terms in the system-bath coupling, beyond the rotating-wave approximation. Furthermore, by computing the spectral function of the quantum emitter we comment on the role played by atom-photon bound-state and show that this acquires a much larger lifetime once the rotating-wave approximation is relaxed.
\end{abstract}

\maketitle

\section{Introduction}


An atom coupled to its electromagnetic environment is one of the most fundamental example of open quantum system and the understanding of spontaneous emission has triggered fundamental advances in this field~\cite{WeisskopfWigner,breuerPetruccione2007}. Since then, the idea of controlling the electromagnetic environment of real or artificial atoms in order to modify and affect their dissipative dynamics has been a much explored and successful one. It has lead for example to the development of cavity~\cite{raimond2001manipulating} and circuit QED~\cite{blais2021circuit}, where the atom is coupled to a resonant mode of a cavity or superconductor resonator. In more recent years the progress in controlling confinement of light in different types of nanostructures and platforms has made the design of photonic environment  a practical experimental tool to modify and shape light-matter interactions between  quantum emitters and electromagnetic modes, and to increase their coupling in the quantum regime.  Examples include arrays of coupled cavities realizing photonic crystals~\cite{Liu2017Quantum}, waveguide coupled to a Quantum Emitter (QE) in Waveguide-QED~\cite{sheremet2021waveguide} or cold atoms arrays~\cite{goban2014atom,hood2016atom,krinner2018spontaneous,stewart2020dynamics}.

Earlier investigations on QEs coupled to structured photonic environments such as photonic crystals displaying a photonic band gap revealed the emergence of atom-radiation bound states, freezing of spontaneous emission and entanglement between matter and light~\cite{john1990quantum,john1994spontaneous,
mogilevtsev2005inreservoir,mogilevtsev2008effective,Lambropoulos_2000}. Recently this topics has received renovated attention~\cite{shi2016bound,gonzalez-tudelaCirac2017,gonzalez-tudelaCirac2017a}. 
Much of these investigations, however, have focused on the weak light-matter coupling regime, where the Rotating-Wave Approximation (RWA), a
conventional approximation in quantum optics, is valid. 
In the regime of strong or ultrastrong light-matter coupling~\cite{kockum2019ultra,forndiaz2019ultra} this approximation can break down and novel phenomena are expected to emerge~\cite{Sanchez_Burillo_2014,diazcamacho2016dynamical,
Rom_n_Roche_2020,Ashida2022Nonperturbative,terradas2022ultrastrong}: in this work we focus on such regimes.

The theoretical description of small quantum systems strongly coupled to structured photonic environments in the non-perturbative light-matter coupling regime poses a number of challenges. 
One cannot make standard Markovian approximations leading to well known master equations, as these rely on the bath spectral function being smooth, which is not the case in presence of band-gaps with sharp edges or Van-Hove types of singularities. 
Also, one needs approaches able to capture the physics in non-perturbative regimes of the light-matter coupling.
For these reasons, in this work we use a recently developed approach for open-quantum systems based on a self-consistent dynamical map~\cite{scarlatella2021noncrossing}.  This approach, based on the self-consistent resummation of infinite class of diagrams in the system-bath coupling corresponding to the so called Non-Crossing Approximation (NCA), allows to treat both the non-perturbative regime of light-matter coupling and to consider a bath density of states which is not necessarily smooth. 


In this work we consider a model for a QE, described as a two-level system, coupled to a two-dimensional array of photonic cavities realizing a photonic crystal with finite bandwidth and sharp band-edges. 
Previous works have already studied this system within the RWA: when the emitter frequency lies outside the band~\cite{john1990quantum} the dynamics is mainly dominated by bound states that are coherent states with a long relaxation time. When the emitter frequency lies inside the band the singularity in the center of the band leads to a non-perturbative regime where the Fermi golden rule fails \cite{Gonz_lez_Tudela_2017_short,Gonz_lez_Tudela_2017}.
We investigate the system beyond the RWA, computing its dynamics 
with the NCA dynamical map both without and within the RWA. 
We consider the cases of an emitter frequency well within the photonic band, outside of the band and at the band edge and we show that in the latter case deviations from the RWA are most pronounced.
We show that several characteristic features of the dynamics near a photonic band-edge, such as the slow-down of the spontaneous decay or the light-matter entanglement are strongly modified in presence of counter-rotating terms. Furthermore, by computing the spectral function of the QE we comment on the role played by atom-photon bound-state and show that this acquires a much larger lifetime once the RWA is relaxed.

This paper is organized as follows. In Sec.~\ref{sec:model} we introduce our model for a quantum emitter coupled to a photonic bath, write down the Hamiltonian for the full light-matter coupling and introduce the RWA to which we will compare our results. In Sec.~\ref{sec:NCA} we discuss the method we use to solve for the dynamics of the QE, namely the NCA dynamical map. In Sec.~\ref{sec:results} we present our results for the dynamics of spontaneous emission, entanglement entropy and for the emitter spectral function. In Sec.~\ref{sec:discussion}, we discuss a qualitative picture to understand the results obtained in the previous section. Finally, In Sec.~\ref{sec:conclusion}  we draw our conclusions. Appendixes contain more details about technical aspects of our work.
%

\begin{figure}[!t] 
    \center 
    \includegraphics[width=0.7\linewidth]{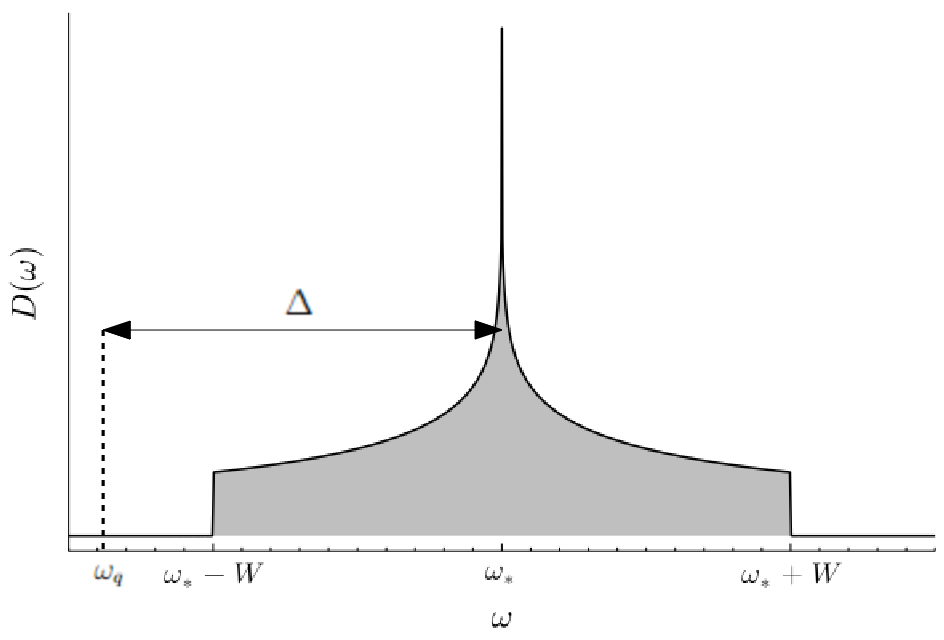}
    \caption{Density of states for the photonic bath realised by a two-dimensional lattice of coupled cavities with nearest-neighbor hopping $J$, as described in the main text. The spectrum is centered around the frequency $\omega_*$ and has a finite half-bandwidth $W=4J$. We note the Van Hove singularity at $\omega_*$ and the sharp band edges at $\omega=\omega_*\pm W$. The quantum emitter frequency is $\omega_q$ and we parametrize it in terms of its distance from the central frequency $\omega_*$, through the parameter $\Delta=\omega_q-\omega_*$. In this respect, an emitter resonant with the bath has $\vert\Delta\vert/W<1$, while for $\Delta=\pm W$ the emitter is at the edge of the band.}
    \label{fig1}
\end{figure}

\section{The Model}\label{sec:model}

We consider a two-level system (TLS) $\{\ket{\uparrow},\ket{\downarrow}\}$ coupled to a photonic bath, describing a Quantum Emitter (QE) or a defect embedded in a photonic crystal. The full Hamiltonian $\m{H}$ of the system can be written as
$$
\m{H}=\m{H}_{S}+ \m{H}_B+\m{H}_{SB}
$$
where $\m{H}_{S}= \omega_q \sigma^z$ is the TLS Hamiltonian with transition frequency $\omega_q$, $\m{H}_{SB}$ is the light-matter coupling while $\m{H}_B$ describes the photonic bath that we take to be a two-dimensional photonic lattice described by a tight-binding model with nearest neighbor coupling $J$ and on-site energy $\omega_{*}$. The Hamiltonian for the bath is given by 
$$
\m{H}_B = \underset{\bb{r}}{\sum}  \omega_{*}a^\dagger_{\bb{r}} a_{\bb{r}} - J \underset{<\bb{r},\bb{r'}>}{\sum} \left( a_{\bb{r}}^\dagger a_{\bb{r'}} + H.c\right)\,,
$$
where $\bb{r}=(x,y)$ denotes the position of the photonic mode in the lattice. By going to Fourier space we can diagonalize the Hamiltonian and bring it to the following form
\begin{align}
    \m{H}_B= \sum_\bb{k} \Tilde{\omega}_\bb{k} a_\bb{k}^\dagger a_\bb{k}
\end{align} 
where we have introduced the dispersion $\Tilde{\omega}_\bb{k}= \omega_* +\omega_\bb{k} = \omega_* - 2J\left[\cos\left( k_x \right)+ \cos\left( k_y \right) \right]$ and the Fourier component of the photon operators $a_\bb{k} = \frac{1}{\sqrt{N}}\underset{\bb{r}}{\sum} e^{-i\bb{k}\cdot \bb{r}}a_\bb{r}$ with $k_{x,y} = \frac{\pi}{N}\left(-N,-N+1 , \cdots , N-1 \right)$.   The main interest of this bath consists in its density of states (DoS)
\begin{align}
D(\omega) = \frac{1}{(2\pi)^2}\int \int d\bb{k} \delta \left(\omega - \Tilde{\omega}_{\bb{k}} \right)
\end{align}
illustrated in the Fig.~\ref{fig1} when $N\rightarrow \infty$. First, the bath has a finite bandwidth, i.e. its frequencies  extend between $\Tilde{\omega}_\bb{k} \in\left[\omega_*-W,\omega_*+W \right]$ where we define $W=4J$ and are centered around $\omega_*$. This allows us to discuss different situations depending on the value of the TLS frequency $\omega_q$, as we discuss below. Furthermore, the DoS has strong non-Markovian features, including sharp band-edges and a Van-Hove singularity in the middle of the band. 

The QE is coupled to the photonic bath locally at a given site through the dipole operator of the TLS which is linearly coupled to the electric field, giving rise to a dipole gauge type of Hamiltonian which reads
\begin{align}
    \m{H}_{SB} = \sigma^x \sum_{\bb{k}}g_\bb{k} \left(a_\bb{k} + a_\bb{k}^\dagger\right)
\end{align}
 In the following, we consider only one mode of polarization of light and a $\bb{k}$-independent coupling constant $g_\bb{k}=g$. Relaxing these constraints also gives rise to interesting phenomena \cite{Gonz_lez_Tudela_2018}. Combining all the terms together we obtain an Hamiltonian for the light-matter system of the form
\begin{align}\label{eq:H_full}
    \m{H} = \omega_q \sigma^z + \sum_\bb{k} \Tilde{\omega}_\bb{k} a^\dagger_\bb{k}a_\bb{k} +\sigma^x \sum_{\bb{k}}g_\bb{k} \left(a_\bb{k} + a_\bb{k}^\dagger\right) 
\end{align}

In the following we will be interested in comparing the dynamics generated by Eq.~(\ref{eq:H_full}) with the one obtained under the RWA. 
This is a widely used approximation in quantum optics which is valid when the bath and QE are close to resonance and weakly coupled: in this case, their dynamics is dominated by their bare frequencies and thus the ``counter-rotating'' terms in the light-matter coupling are rapidly oscillating and can be neglected. This yields
\begin{align}
\label{eq:H_rwa}
    \m{H}_{RWA} = \omega_q \sigma^z +  \sum_{\bb{k}} g_{\bb{k}} \left(a_\bb{k}\sigma^+ + a_\bb{k}^\dagger \sigma^- \right) + \sum_{\bb{k}}  \Tilde{\omega}_\bb{k} a_\bb{k}^\dagger a_\bb{k}
\end{align}
In this form the Hamiltonian conserves the total number of excitations $N_{exc}=\sigma^z + \sum_\bb{k} a_\bb{k}^\dagger a_\bb{k}$ and allows for a simple solution in the subspace at fixed $N_{exc}$. In particular, for $N_{exc}=1$ several results have been known in the literature. In the single excitation sector, the total state of the system described by $\m{H}_{RWA}$ can be written explicitly, if we suppose a empty bath initially $\ket{\bb{0}}$.
\begin{align}
    \ket{\psi(t)} = \left( C_{eg}(t)\sigma^+  + \sum_\bb{k} C_{k}(t) a_\bb{k}^\dagger \right) \ket{\downarrow}\otimes \ket{\bb{0}}
\end{align}
using this expression and the propagator we can diagonalize the hamiltonian $\m{H}_{RWA}$ and also obtain the dynamics of spontaneous emission~\cite{Gonz_lez_Tudela_2017_short}. Note that when the band admits discontinuities or singularities, it is necessary to carry out analytical continuations or approximations in order to obtain the dynamics.
Here we will discuss the effects of counter-rotating terms on the physics of this model, as a function of the light-matter coupling $g$ and the frequency of the emitter $\omega_q$ that we parametrize in terms of $\Delta=\omega_q -\omega_*$, the detuning with respect to the middle of the band $\omega_{*}$ (See Fig.~\ref{fig1}). This formally corresponds to go to a frame rotating at the central frequency of the photonic band $\omega_*$, by applying the unitary transformation $\hat{U} = \exp\left(it \omega_* \left( \sigma^z +\sum_\bb{k} a_\bb{k}^\dagger a_\bb{k}\right) \right)$. We note that in the limit $\omega_*, \omega_q\rightarrow\infty$ we expect the effect of counter-rotating terms to disappear and the RWA to become exact.

\section{NCA Dynamical Map}\label{sec:NCA}

\begin{figure}[t]
    \center 
    \includegraphics[width=0.8\linewidth]{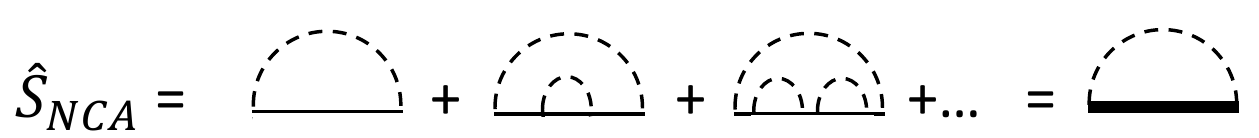}
    \caption{Self-energy in the non-crossing approximation (NCA). Solid
lines correspond to bare time evolution superoperators (without the bath), dashed lines correspond to bath 2-times correlation functions, bold solid lines correspond to the dynamical map $\m{V}$. Only diagrams where the dashed lines do
not cross are kept in the non-crossing approximation.}    
\label{fig:schema_NCA}
\end{figure}

In this section we briefly discuss the theoretical approach we use to compute the dynamics of our model. Since we are interested to go beyond the weak-coupling regime in which RWA is valid, we resort to the recently developed Non-Crossing dynamical map~\cite{scarlatella2021noncrossing}. 
This approach is similar in spirit to a master equation, describing the reduced dynamics of the QE after the bath is integrated out, but it yields an equation for the dynamical map of the system, that is the superoperator $\hat{\m{V}}(t)$ evolving the reduced density matrix, rather than for the density matrix itself. 
The dynamical map is defined by $\rho(t) =Tr_B \rho_{tot}(t) = \hat{\m{V}}(t) \rho(0)$; expanding in the system-bath coupling, one finds that it obeys the Dyson equation
\begin{align}
\partial_t \hat{\m{V}}(t) \bullet = -i \left[ H_S, \hat{\m{V}}(t) \bullet \right] + \int_{0}^t \Sigma(t-t_1) \hat{\m{V}}(t_1) \bullet
\end{align}
where $\Sigma$ is defined as an infinite series involving the 2-times correlations functions of the bath and where the bullets indicate the arguments of super-operators, when necessary.
Within a non-crossing approximation, described in Fig. \ref{fig:schema_NCA}, and without making the RWA, the self-energy takes the simple analytical expression
\begin{equation}
\label{eq:ncaSelfEn}
\Sigma_{\rm NCA}(\tau) = \Gamma(\tau) \left(  \hat{\m{V}}(\tau) \left[ \sigma^x \bullet  \right] \sigma^x    -   \sigma^x \hat{\m{V}}(\tau) \sigma^x \bullet \right) + \rm{h.c.} 
\end{equation}
$\Gamma(\tau)$ is the 2-times correlation function of the bath operator $
    \Gamma(\tau) = \Tr_B\left[B(\tau) B(0) \rho_B(0) \right]
 $ 
with $B=\sum_{\bb{k}}g_\bb{k}\left( a_\bb{k} + a_\bb{k}^\dagger\right)$ and $a_\bb{k}^{(\dagger)}(\tau)$ the creation (annihilation) bosonic operator in the interaction picture.
We note that the self-energy has a similar structure to the ``dissipator'' of standard master equations, and in fact the Born Master equation can be recovered by replacing $\m{V}(\tau)$ by $e^{\m{H}_S \tau}$ in the expression \eqref{eq:ncaSelfEn} of the self energy.
The expression for the self-energy within the RWA is reported in the Appendix~\ref{app:dyson}.

\section{Results}\label{sec:results}

In this section, we discuss our results for the dynamics of the system obtained with the NCA dynamical map. We consider the emitter initially in the excited state $\rho \left(0\right)=\ket{\uparrow}\bra{\uparrow}$
coupled to an initially empty bath, described by $\rho_B(0) = \ket{\bb{0}} \bra{\bb{0}}$. Then we let
the system evolve through the Hamiltonian with \eqref{eq:H_rwa} and without RWA \eqref{eq:H_full}, and study the emitter relaxation.


We fix $\omega_*\simeq W$ and we voluntarily choose a large bandwidth, in order to minimize the influence of the edge (singularity in the middle) of the band when the emitter frequency lies in the middle of the band (edge).



\subsection{Dynamics of Spontaneous Emission from weak to strong coupling}

\begin{figure}[t] 
    \center 
    \includegraphics[width=\linewidth]{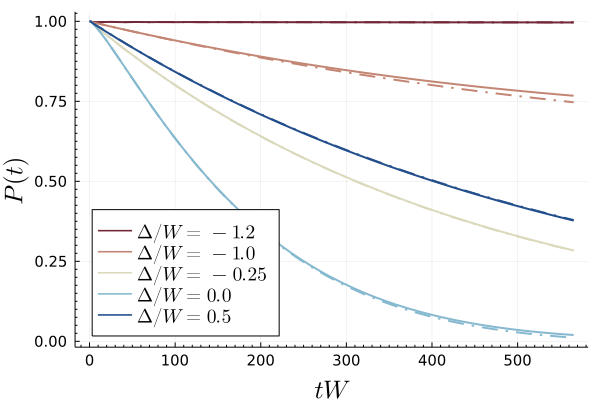}
    \caption{Spontaneous emission of a single quantum emitter, initially prepared in its excited state, and then suddenly coupled to a photonic bath, with a coupling constant $g/W=0.05$ and for
different detuning values $\Delta/W =-1.2,-1,-0.25,0,0.5 $, corresponding to a QE outiside the band ($\Delta/W<-1$) at the edge ($\Delta/W=-1$)  or within the band ($\Delta/W>-1$). Solid lines correspond to the dynamics of $\m{H}$ and that dotted at $\m{H}_{RWA}$. }
    \label{fig:spontadelta}
\end{figure}

We start discussing the dynamics of spontaneous emission, described by the time-dependent population of the TLS
\begin{align}
P(t) \equiv \frac{1}{2}\left( 1+\left<\sigma^z\right>(t) \right)  
\end{align}
In Fig.~\ref{fig:spontadelta}, we show the results obtained numerically for the dynamics of the emitter, for different values of detuning between the emitter and the center of the photonic band $\Delta/W =-1.2,-1.0,-0.25,0,0.5 $  and for a fixed coupling constant $g/W=0.05$. Solid and dotted lines correspond respectively to the dynamics without and within the RWA. 
As expected, we find that the decay of the spontaneous emission is faster when the frequency of the emitter lies within the photonic band, $|\Delta/W|<1$, while it slows down as its frequency is moved outside of the band since the emitter cannot hybridize with the bath. The decay is exponential in time and can be estimated using perturbation theory and Fermi Golden's Rule (FGR) to give $\Gamma_{FGR}(\Delta) = 2\pi g^2 D(\Delta)$. This is true except if the emitter frequency is close enough to the middle of the band, where the density of states features a Van-Hove singularity: in this case the decay is non-exponential and perturbation theory breaks down as discussed in the literature  \cite{yuanYi2017,Gonz_lez_Tudela_2017_short,Gonz_lez_Tudela_2017}.
%

We see from Fig.~\ref{fig:spontadelta} that for (almost) all detunings, including when the emitter is resonant with the singularity $\Delta/W=0$, the dynamics within and without the RWA agrees for the (weak) light-matter coupling considered. 
We also remark (not shown) that the NCA dynamical map is able to predict a deviation from exponential relaxation when the emitter frequency is sufficiently close to the singularity. 
Particularly interesting for this work is the regime in which the emitter frequency is at the lower edge of the photonic band, corresponding to $\Delta/W=-1$ in Fig.~\ref{fig:spontadelta}: in this case we see that significant deviations arise depending on whether one makes the RWA or not.
This observation suggests that going beyond the RWA is necessary to describe the physics in this regime, due to the sharp discontinuity of the bath DoS at the edge of the band. 


%

\begin{figure}[t] 
    \center 
    \includegraphics[width=1.0\linewidth]{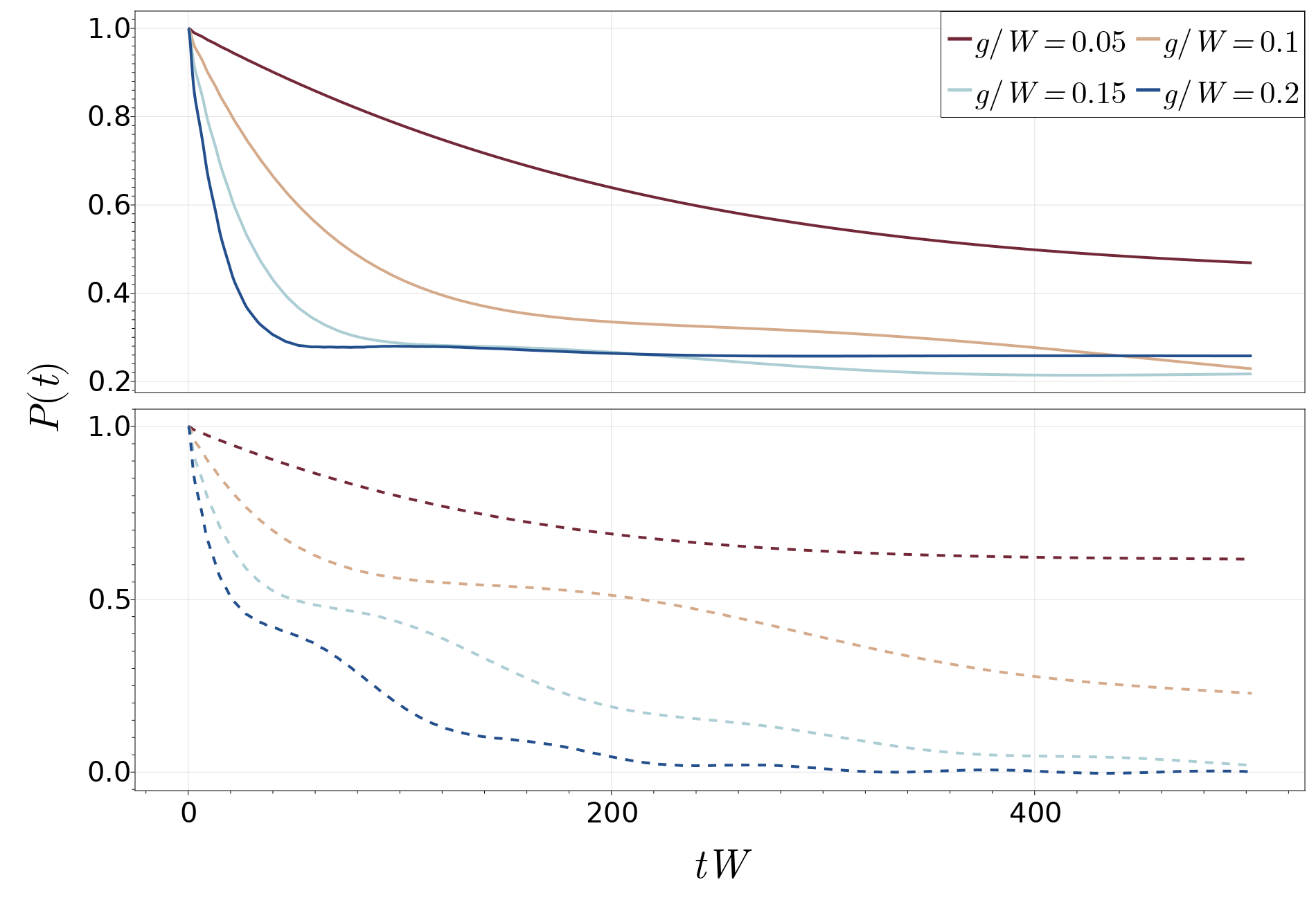}
    \caption{Spontaneous emission of a single quantum emitter with frequency at the lower edge of the photonic band, corresponding to $\Delta/W=-1$, initially prepared in its excited state and then suddenly coupled to the bath. The top panel correspond to the dynamic of the full Hamiltonian while the bottom panel is the evolution of the entanglement entropy for the  Hamiltonian under RWA.}
     \label{fig:spontag}
\end{figure}

In Fig.~\ref{fig:spontag} we focus on the regime in which the emitter frequency is at the edge of the photonic band $\Delta/W=-1$, where we expect the deviations from RWA to be most important, and vary the light-matter coupling $g$ from weak to strong. 
The top panel corresponds to the evolution obtained within the RWA and the bottom panel to the dynamics generated by the full Hamiltonian.


In this latter case we see that upon increasing $g$ the exponential relaxation of the spontaneous decay observed before is modified. In particular we observe the expected freezing of spontaneous emission which appears in the form of intermediate time plateau in the time evolution. For $g/W=0.05$ for example we clearly see a rapid decay of $P(t)$ followed by an almost constant evolution up to some longer time scales at which the dynamics relaxes again towards zero. As the coupling is further increased the metastable plateau is reduced, however we see that at stronger couplings multiple plateau re-emerge. This is a phenomenology that is strongly reminiscent of prethermalization in weakly non-integrable many-body systems which possess several well separated energy scales corresponding to the unlocking of almost conserved quantities. 
This phenomenology is modified by the presence of counter-rotating terms (top panel) which give rises to important differences in the dynamics of the spontaneous emission. In particular we note a much faster decay towards the steady state, which quite importantly does not need to be the ground-state of the Qubit since counter-rotating terms can support a non trivial steady state. Quite interestingly, we see that the plateau structure reminiscent of metastability is not completely washed away, as we see at intermediate and strong coupling $g/W=0.1-0.2$.

Within the RWA the freezing of spontaneous emission is usually understood with the appearance of coherent state in the dynamics, which live outside the band (bound state) \cite{mogilevtsev2005inreservoir,D_az_Camacho_2016,Rom_n_Roche_2020,Sanchez_Burillo_2014}. At the lower edge of the band, the frequency of the emitter splits between coherent states in the gap and modes in the band \cite{john1990quantum}, is it in this that we find a dynamic which for a long time relaxes towards the ground state of the QE. Here, for the full dynamics this picture will need to be modified as we will discuss in Sec.~\ref{sec:discussion}.

\begin{figure}[t] 
    \center 
    \includegraphics[width=\linewidth]{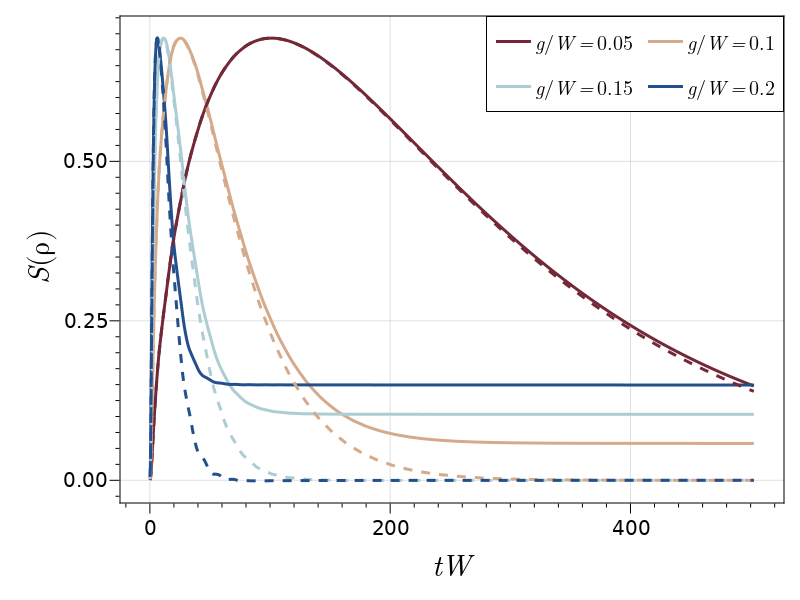}
    \caption{Dynamics of the entanglement entropy for a QE with detuning within the photon band,  $\Delta/W=-0.75$ and for different values of coupling $g/W$ with the bath. Solid lines correspond to the dynamics of $\m{H}$, and the dotted one the dynamics of $\m{H}_{RWA}$. }
     \label{fig:ee_intheband}
\end{figure}
\subsection{Dynamics of Entanglement Entropy}

To better understand the dynamics at strong coupling and the role of counter-rotating terms we now discuss the evolution of the entanglement entropy between the QE and its photonic environment. We emphasize that the full system plus environment evolve unitarily and therefore we can consider a bipartition containing the emitter and define its reduced density matrix as $\rho(t)=\Tr_B\rho_{tot}(t)$, where the trace is taken over the photonic bath degrees of freedom. Then the entanglement entropy between emitter and its environment reads
\begin{align}   
 S(t) = - \Tr \left( \rho(t) \log \rho(t)\right)\,.
\end{align}
We note incidentally that the emitter reduced density matrix  $\rho(t)$ is a quantity  we can naturally access with the NCA dynamical map. For spin-boson types of models the study of entanglement entropies in equilibrium and at the quantum phase transition has received major attention in the past~\cite{kopp2007universal,lehur2007entanglement,HUR20082208}.

We first consider the case in which the QE lies within the photonic band, $\Delta/W=-0.75$, and different values of system-bath coupling $g/W$ and compare in Fig.~(\ref{fig:ee_intheband}) the full dynamics obtained within NCA with the RWA. We see that in both cases the entanglement grows with time, starting from zero since the initial state we consider is a product state between system and bath,  reaches a maximum value and then decreases. The value of this maximum entanglement does not depend on the coupling $g$ and concides with the maximum entanglement for a two-level system, i.e. the system is maximally entangled at short times due to the coupling to the photonic bath. On the other hand, the time scales to reach the maximum entanglement and then to leave it towards the stationary value  depend strongly on $g$: we see that quite generically the dynamics is much faster as $g$ is increased. The comparison between full dynamics and RWA reveals that the dynamics of entanglement is similar, while the long-time limit is different. In particular the counter-rotating terms are able to sustain a finite entanglement entropy at long times, while within RWA the entanglement always goes to zero at long times since the QE relaxes back to the ground state.
%


\begin{figure}[t] 
    \center 
    \includegraphics[width=1.0\linewidth]{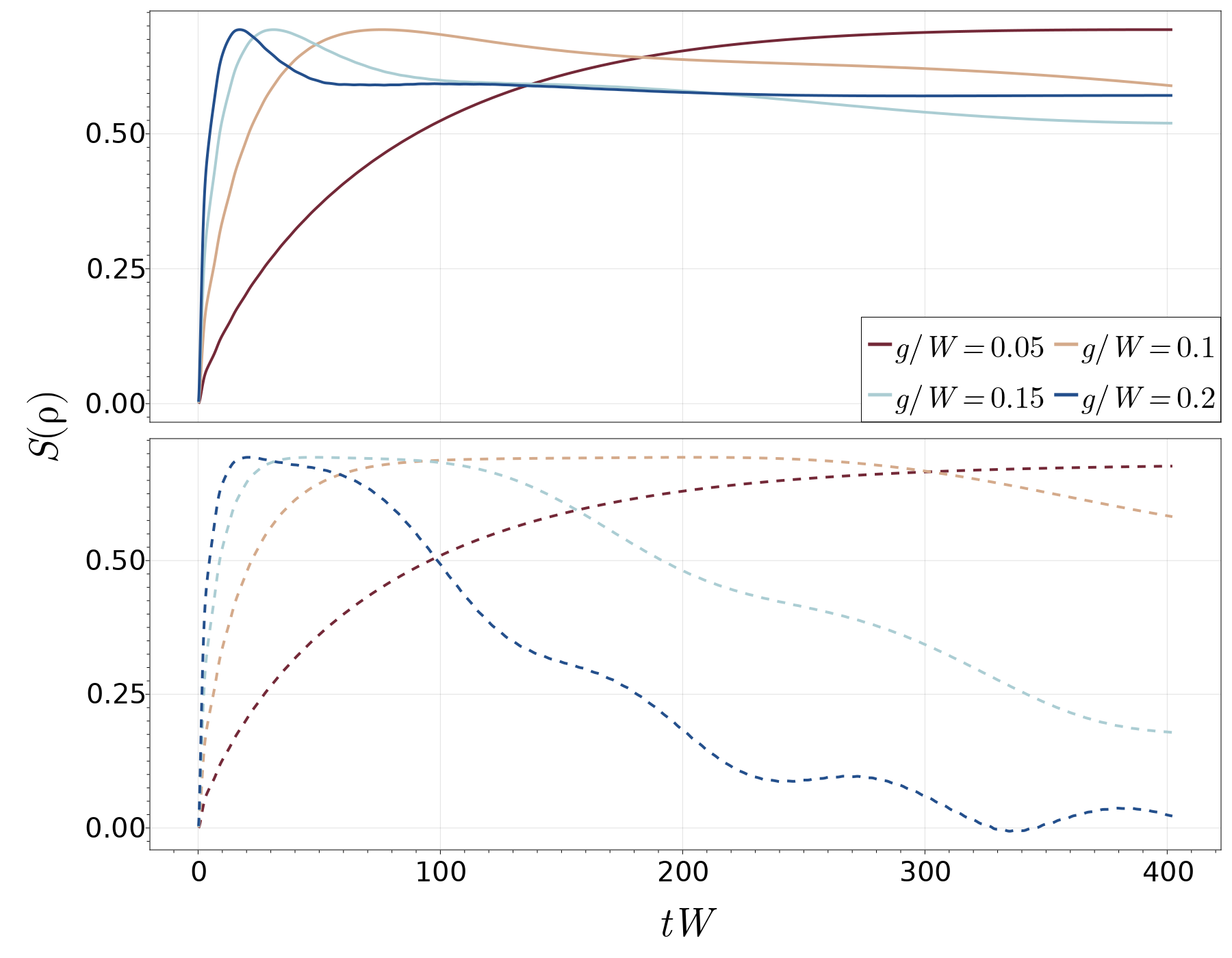}
    \caption{Dynamics of the entanglement entropy for a QE with detuning at the edge of the photonic band, $\Delta/W=-1$ and for different values of coupling $g/W$ with the bath. The top panel correspond to the dynamic of the full Hamiltonian while the bottom panel is the evolution of the entanglement entropy for the  Hamiltonian under RWA.}
     \label{fig:ee_edge}
\end{figure}

We now discuss the more relevant case for us of a QE sitting at the edge of the photonic band, i.e. $\Delta/W=-1$. In Fig.~\ref{fig:ee_edge} we plot the resulting dynamics of the entanglement entropy for the full Hamiltonian (top panel) and the RWA case (bottom panel). Focusing on the latter case first we note that the main difference that appears with respect to Fig.~\ref{fig:ee_intheband} is the fact that the maximum entanglement is stabilized and mantained for a finite time-interval, before decaying to zero. This maximally entangled time interval depends on the system-bath coupling, i.e. for weaker coupling the plateau is longer-lived while it is destroyed rapidly for strong $g$. This phenomenon,  which corresponds to a back action of the reservoir between the emitter and the mode of the photonic band \cite{john1994spontaneous,kilin1992freezing,1993OptSp74579K,john1995localization} is linked to bounds state in the dynamics.  It can be understood as the entanglement entropy counterpart of the prethermalization plateau observed in the dynamics of spontaneous emission discussed before. In particular we can use the same kinematic argument and the conservation of total excitation number to understand the resilience of entanglement to decay in the case of RWA dynamics.
 This picture is further confirmed by looking at the full dynamics beyond RWA, shown in the top panel. There we see that the maximum entanglement is preserved at short times: the rapid increase of entanglement entropy seems not affected by the counter-rotating terms which instead play a more relevant role at longer times, destabilizing the maximally entangled state on much shorter time scales and leading to a stationary state which is not the trivial ground-state of the isolated emitter but contains non-zero system-bath entanglement. 
%

\begin{figure}[t] 
    \center 
    \includegraphics[width=1.0\linewidth]{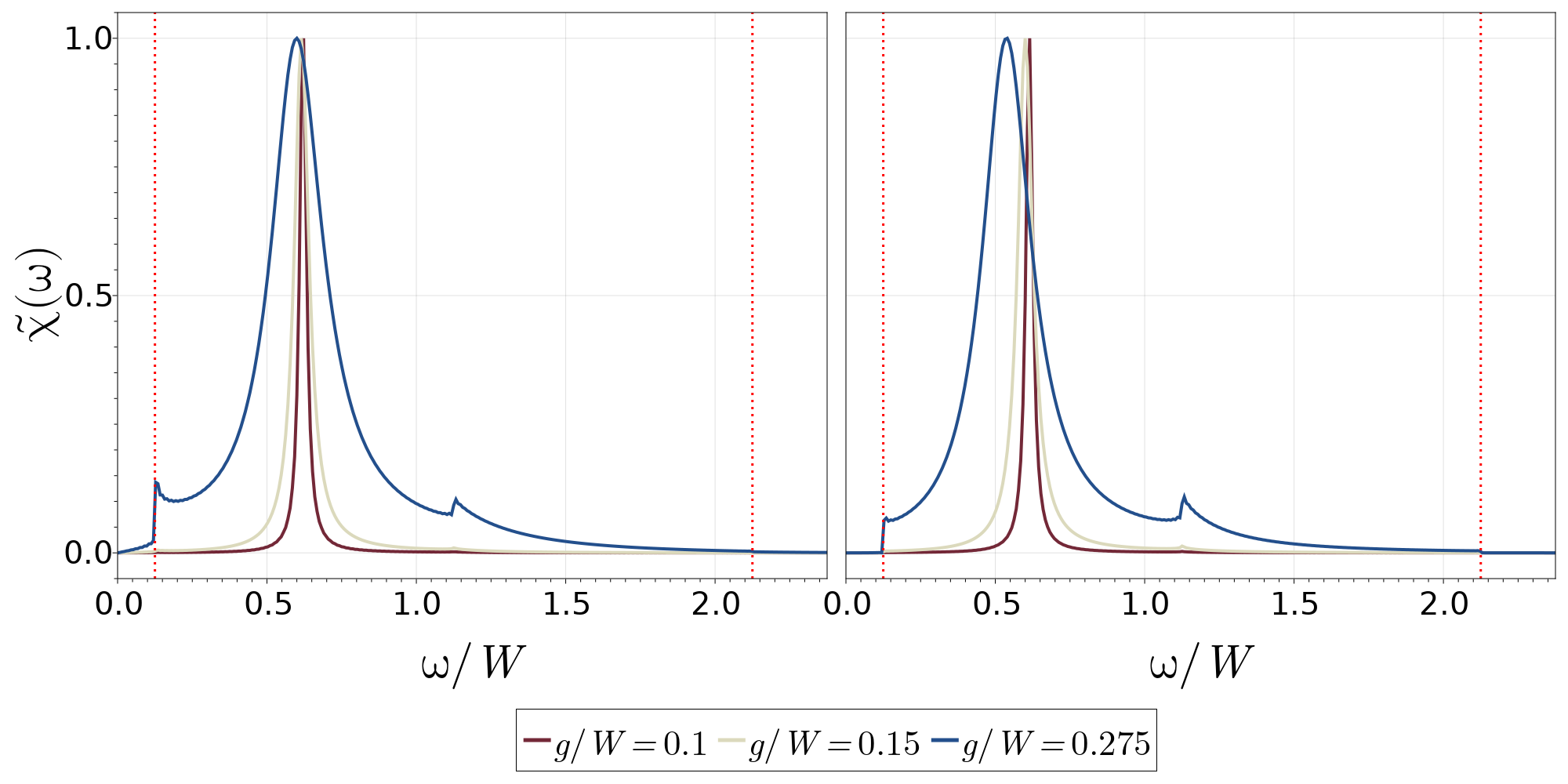}
    \caption{Spectral function of the QE, as defined in the text, for differents values of coupling $g/W$ and for a QE frequency lying within the photonic band, $\Delta/W=-0.5$, corresponding to a frequency $\omega_q/W=0.625$ We plot only positive frequencies and represent with the dotted lines the  upper and lower edges of the photonic bath. Left panel shows the result for the full Hamiltonian, while the right panel within the RWA.  For weak couplings we see in both cases a a narrow resonance centered around $\omega=\omega_q$ and a small feature in correspondence of the singularity in the middle of the band. For stronger couplings the deviations between full Hamiltonian and RWA become more evident, both in terms of width of the resonance and behavior at the edge of the band.} \label{fig:spectrum}
\end{figure}

\subsection{Photonic Bound-State and Quantum Emitter Green's Function}

A typical approach to understanding the link between dynamics and the nature of excitations is
to look at the spectral function of the system encoded in the retarded Green's function of the emitter,
which is defined as
\begin{align}\label{eqn:chi_t}
    \chi(t) = -i\theta(t)\left<\left[\sigma^x(t),\sigma^x(0) \right] \right>\,.
\end{align}
Here, the average is taken over the stationary density matrix of the coupled system and bath and the time-evolution is performed with respect to the full Hamiltonian of system and bath.  This quantity can be computed within our NCA dynamical map, by expressing Eq.~(\ref{eqn:chi_t}) in terms of the evolution superoperator $\hat{\m{V}}(t)$ defined in Sec.~\ref{sec:NCA}~\cite{scarlatella2021noncrossing}. From this quantity, we can extract the Fourier transform and take the imaginary part which contains information about the spectrum of the system, i.e. $\chi(\omega)=\mbox{Im}\int dt e^{-i\omega t}\chi(t)$.

We start considering the case in which the emitter frequency lies within the photonic band, i.e. $\Delta/W =-0.5$ and plot in Fig.~\ref{fig:spectrum}
the spectral function for different values of light-matter coupling $g$, both in the case of RWA (right panel) and for the full Hamiltonian (left panel). 
In both cases we see some common features emerging in the spectral function, including  a peak within the band (dotted lines represent the upper and lower edges of the band) corresponding to the frequency of the emitter $\omega_q/W=0.625$ which shifts and  becomes broader as the coupling $g$ is increased. At stronger values of the coupling we see also the appearance (in both panels) of a spectral feature in the middle of the band corresponding to the Van-Hove singularity. On the other hand, we see that the behavior of the spectral function at frequencies near the band-edge is rather different and that the presence of counter-rotating terms in the full Hamiltonian has direct consequences on the spectral features of the system.  In particular as we increase the coupling $g$ we see that states in between the gap appears in the spectrum, corresponding to processes involving virtual photons that cannot appear within RWA due to the conservation of total number of excitations. Furthermore, we see the emergence of a sharp peak at the edge of the band which is completely absent in the RWA data and which gets stronger as $g$ increases. We will comment on the origin of this spectral feature in Sec.~\ref{sec:discussion}.
%


\begin{figure}[t] 
    \center 
    \includegraphics[width=0.9\linewidth]{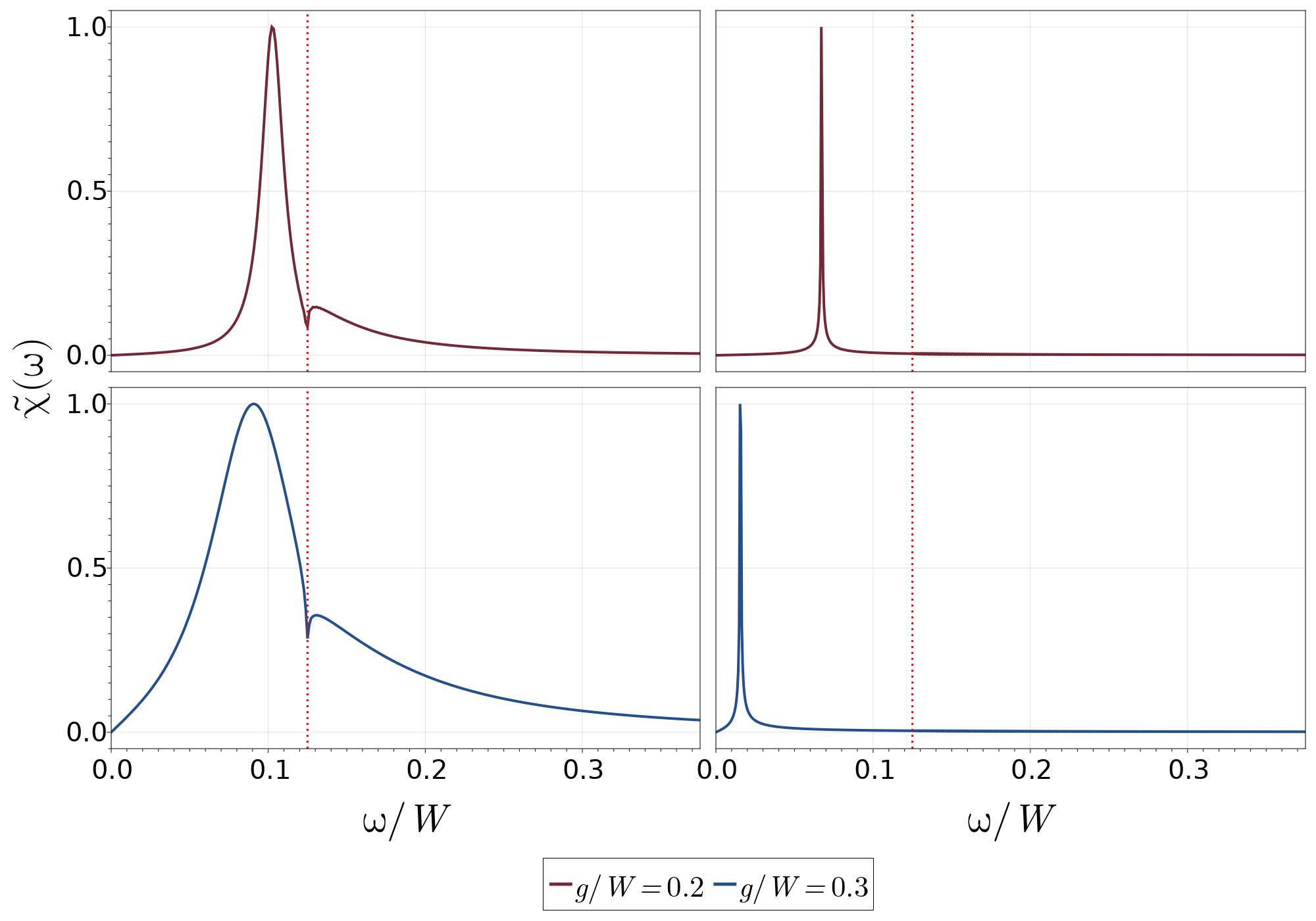}
    \caption{
Spectral function of the QE, as defined in the text, for differents values of coupling $g/W$ and for a QE frequency lying at the edge of the photonic band, $\Delta/W=-1$. We plot only positive frequencies and represent with the dotted lines the  upper and lower edges of the photonic bath. Left panel shows the result for the full Hamiltonian, while the right panel within the RWA. We see that the deviations appear already for moderately weak couplings, where the RWA results show a narrow resonance within the gap, corresponding to a photonic bound-state, while the inclusion of counter-rotating terms leads to a broadening of the in-gap peak and a transfer of spectral weight within the band.}
%
%
\label{fig:spectrum2}
\end{figure}

We now discuss the spectral function for a case in which the emitter frequency lies at the edge of the band, which for the dynamics corresponds to the stronger differences between RWA and full dynamics. In Fig.~\ref{fig:spectrum2} we plot the spectral function for $\Delta/W=-1$ and different values of the coupling $g$ respectively for the full Hamiltonian (left panel) and the RWA case (right panel). We see that in this case major differences appear in the spectrum already for relatively small values of light-matter coupling. In particular within the RWA there is a narrow peak within the photonic band gap whose position depends on $g$ and which corresponds to a photonic bound-state. Once counter-rotating terms are included they lead to two major effects, namely a broadening of the in-gap peak which becomes a resonance and acquires a finite-lifetime and a transfer of spectral weight into the photonic bath. We also notice the presence of a features right at the band-edge. In both cases, we observe a splitting of the emitter excitation into a coherent excitations outside the band and others modes living in the band. This splitting corresponds to the freezing of the spontaneous emission, indeed the band modes have a faster decay time than the coherent excitations outside the band.  Moreover, this splitting is all the more important and brings into play all the more modes
that the coupling with the bath is strong.

\section{Discussion}\label{sec:discussion}

\begin{figure}[t]
    \center 
    \includegraphics[width=0.8\linewidth]{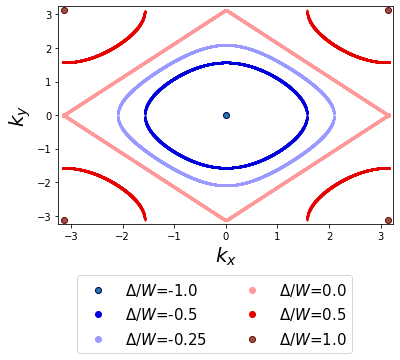}
    \caption{Graphical representation of the modes  resonant with the frequency of the Quantum Emitter, $\omega_\bb{k}=\Delta$. We notice a continuum of resonant modes when the quantum emitter frequency is located in the band, $\vert\Delta\vert/W<1$, which shrink to a point as $\Delta$ approaches the lower edge of the photonic band, $\Delta=-W$. On the other hand, at the upper edge $\Delta=W$ the resonant condition is satisfied by four state.}    
\label{fig:schema_Discussion}
\end{figure}

The results shown in the previous section highlight the fact that counter-rotating terms in the system-bath Hamiltonian become relevant for the dynamics of the system not only at strong coupling $g$, but also at weak and intermediate values provided the frequency of the quantum emitter is resonant with the sharp band edge. In order to qualitative understand these results it is useful to perform a time-dependent unitary transformation and to rewrite the full Hamiltonian of $H$ in the interaction picture with respect to the free evolution of emitter and bath, $H_0=\omega_q \sigma^z +\sum_\bb{k} \tilde{\omega}_\bb{k}a_\bb{k}^\dagger a_\bb{k}$. This gives
\begin{align}\label{eqn:H_rot}
    \m{H}(t)& =g\sum_\bb{k}\left(e^{i\left(\omega_q- \Tilde{\omega}_\bb{k}\right) t} \sigma^+ a_\bb{k}+ \mbox{h.c.}  \right)+\nonumber\\
&+    g\sum_\bb{k}\left(e^{i\left(\omega_q+ \Tilde{\omega}_\bb{k}\right) t} \sigma^+ a_\bb{k}^\dagger + \mbox{h.c.} \right)
\end{align}
where the system-bath coupling is now explicitly time-dependent and contains two types of terms, those conserving the total number of excitations and oscillating at frequency $\omega_q- \Tilde{\omega}_\bb{k}$ and the counter-rotating terms oscillating at frequency $\omega_q+ \Tilde{\omega}_\bb{k}$.  The RWA amounts to disregard the latter terms which are rapidly oscillating as compared to the number-conserving couplings. This approximation is usually valid provided that 
$$
\vert\omega_q+ \Tilde{\omega}_\bb{k}\vert \gg \vert \omega_q- \Tilde{\omega}_\bb{k}\vert 
$$
 This is true in particular whenever the emitter frequency is resonant with the photonic band, $\omega_q=\Tilde{\omega}_\bb{k}$ or equivalently $\Delta=\omega_\bb{k}$, which is the case (for some value of $\bb{k}$) whenever $\vert \Delta \vert/W<1$ (see Fig.~\ref{fig:schema_Discussion}).
 
 We expect therefore the RWA to work well whenever the emitter is resonant with many modes of the photonic bath. When this is the case the Hamiltonian in Eq.~(\ref{eqn:H_rot}) can be split into a sub-set of $\bb{k}$ modes which are resonant with the emitter and static, their counter-rotating terms oscillating at frequency $2\omega_q$ acting as a drive and a set of complementary bath modes which are off-resonant and oscillate with multiple frequencies and include both rotating and counter-rotating terms.  The condition $\Delta=\omega_\bb{k}$ is met by an extensive number of $\bb{k}=0$ points if, for example, $\Delta=0$ in the middle of the band, so the subset of coherently coupled resonant modes act itself as a bath and provides dissipation. In this regime one expect a fast dynamics of the emitter and a good agreement between RWA and the full dynamics, except on long time scales where the counter-rotating terms can stabilize a non-trivial stationary state. However, as $\Delta$ approaches the lower edge of the photonic band the number of resonant $\bb{k}$ modes decreases. Right for $\Delta=-W$, corresponding to the emitter at the edge of the photonic band, this condition is only met at $k_x=k_y=0$. This suggests an effective Jaynes-Cumming or Rabi type of model in which the emitter is coherently coupled to the resonant $\bb{k}=0$ mode, with time-dependent coherent drive of counter-rotating terms and off-resonantly coupled to the rest of the bath modes which provide dissipation. 
 \begin{align}\label{eqn:H_rot2}
    \m{H}_{\rm eff}(t)& =g \left(\sigma^+ a+ \mbox{h.c.}  \right)+g \left(\sigma^+ a^{\dagger}e^{2i\omega_q t}+ \mbox{h.c.}  \right)+
\notag \\&+     g\sum_\bb{k}\left(e^{i\left(\omega_q- \Tilde{\omega}_\bb{k}\right) t} \sigma^+ a_\bb{k}+ \mbox{h.c.}  \right)+\notag\\&
+g\sum_\bb{k}\left(e^{i\left(\omega_q+ \Tilde{\omega}_\bb{k}\right) t} \sigma^+ a_\bb{k}^\dagger + \mbox{h.c.} \right)
\end{align}
 
 Furthermore all the couplings in the RWA have to conserve the total number of excitations. In this picture the freezing of spontaneous emission and the entanglement plateaux arise because of the interplay between off resonant system-bath couplings, which are responsible for bringing the system back to its ground-state but are kinematically blocked at small $g$ because of the conservation of excitation number, and resonant coherent coupling with the  $\bb{k}=0$ mode which instead protects entanglement and a finite value of the emitter polarization. On the other hand when the counter-rotating terms are included the picture changes substantially: from one side new coherent coupling terms arise between the emitter and the $\bb{k}=0$ mode  and new off resonant dissipative channels emerge, both of which break the conservation of excitations and lead to a non-trivial stationary state but also spoils the separation of energy scales associated to the metastable states observed in the RWA.

The argument above can also be used to understand qualitatively the origin of the strong spectral response of the system at frequency corresponding to the lower edge of the band, as shown in Fig.~\ref{fig:spectrum}. In fact we note that in the rotating frame picture discussed above the full Hamiltonian can be obtained from the RWA one by including a time-dependent perturbation given by the counter-rotating terms, which oscillate at
$\omega_q + \tilde{\omega_\bb{k}}$. In the high-frequency and weak coupling limit, corresponding to the regime of validity of RWA as discussed before , we expect this perturbation to be irrelevant for the physics of the system. However we note that the modes at the edge of the band  correspond to the frequencies $\bb{k}=0$ 
and are always the slowest ones $\omega_q + \tilde{\omega_\bb{k}}$. For this reason we expect that in the response to a weak drive at that frequency, which corresponds essentially to the spectral function discussed in Sec.~\ref{sec:results}, the deviations between RWA and full dynamics to be more visible at frequencies near the lower band edge. 

Finally, we note that while the validity of RWA is usually assumed for resonant couplings, we can provide an extended criterion of validity of RWA, which reads, in the case the initial density matrix contains a single excitation (See Appendix~\ref{app:validity}),
\begin{align}
    g^2 \underset{\tilde{\omega}_{\bb{k}_1},\tilde{\omega}_{\bb{k}_2}}{max}{\left[\frac{1}{\left(\tilde{\omega}_{\bb{k}_1}+\tilde{\omega}_{\bb{k}_2}\right) \left(\omega_q + \tilde{\omega}_{\bb{k}_1} \right)}  \right]} \ll 1
\end{align}
and that gives the condition 
\begin{align}
    g^2 \ll \left( 2\omega_* - 8J\right) \left(2\omega_*+\Delta - 4J \right)
\end{align}
explaining why the RWA fails more strongly when the emitter is at the lowest edge of the photonic band.


\section{Conclusion}\label{sec:conclusion}

In this work we have studied the dynamics of a quantum emitter strongly coupled to a photonic environment featuring a finite band-widht and a band gap. We have discussed in particular the role of the counter-rotating terms in the full Hamiltonian and the consequences they bring with respect to the dynamics within the Rotating Wave Approximation (RWA). To solve the resulting spin-boson model we have used the recently developed NCA dynamical map, which gives direct access to the reduced density matrix of the emitter, from which local properties as well as spontaneous emission, entanglement entropy can be readily obtained as well as the frequency resolved spectral function.
We have shown that the major deviations from RWA due to the counter-rotating terms arise when the emitter frequency is at the lower edge of the photonic band where the freezing of spontaneous emission and the maximum entanglement, usually interpretred in terms of photonic bound-states, are destroyed and the system is driven towards a non-trivial and entangled stationary state. We have shown that spectral features corresponding to this physics arise in the emitter spectral function. Specifically we have shown that the narrow in-gap peak appearing in the RWA case becomes a much broader resonance due to counter-rotating terms and also hybridizes with modes within the photonic band. We have provided a qualitative picture to understand these results in terms of an effective Rabi-like model, featuring a resonant coherent coupling between emitter and the $\bb{k}=0$ mode and to a bath of off-resonant excitations which provide dissipation. Our results show the importance of counter-rotating terms in the dynamics of wave-guide QED systems not only at ultrastrong coupling but also at intermediate light-matter coupling regimes, depending on the relative detuning of the emitter. Directions for future work could include for example the interplay of two or more quantum emitters, which can be still treated within the NCA dynamical map, or the calculation of photonic bath properties to inquire the detail structure of the system-bath wave function at strong coupling.

\section*{Acknowledgments} 
This project has received funding from the European Research Council (ERC) under the European Union’s Horizon 2020 research and innovation programme (Grant agreement No. 101002955 — CONQUER).

\bibliographystyle{apsrev4-1}   

%



\begin{thebibliography}{37}%
\makeatletter
\providecommand \@ifxundefined [1]{%
 \@ifx{#1\undefined}
}%
\providecommand \@ifnum [1]{%
 \ifnum #1\expandafter \@firstoftwo
 \else \expandafter \@secondoftwo
 \fi
}%
\providecommand \@ifx [1]{%
 \ifx #1\expandafter \@firstoftwo
 \else \expandafter \@secondoftwo
 \fi
}%
\providecommand \natexlab [1]{#1}%
\providecommand \enquote  [1]{``#1''}%
\providecommand \bibnamefont  [1]{#1}%
\providecommand \bibfnamefont [1]{#1}%
\providecommand \citenamefont [1]{#1}%
\providecommand \href@noop [0]{\@secondoftwo}%
\providecommand \href [0]{\begingroup \@sanitize@url \@href}%
\providecommand \@href[1]{\@@startlink{#1}\@@href}%
\providecommand \@@href[1]{\endgroup#1\@@endlink}%
\providecommand \@sanitize@url [0]{\catcode `\\12\catcode `\$12\catcode
  `\&12\catcode `\#12\catcode `\^12\catcode `\_12\catcode `\%12\relax}%
\providecommand \@@startlink[1]{}%
\providecommand \@@endlink[0]{}%
\providecommand \url  [0]{\begingroup\@sanitize@url \@url }%
\providecommand \@url [1]{\endgroup\@href {#1}{\urlprefix }}%
\providecommand \urlprefix  [0]{URL }%
\providecommand \Eprint [0]{\href }%
\providecommand \doibase [0]{http://dx.doi.org/}%
\providecommand \selectlanguage [0]{\@gobble}%
\providecommand \bibinfo  [0]{\@secondoftwo}%
\providecommand \bibfield  [0]{\@secondoftwo}%
\providecommand \translation [1]{[#1]}%
\providecommand \BibitemOpen [0]{}%
\providecommand \bibitemStop [0]{}%
\providecommand \bibitemNoStop [0]{.\EOS\space}%
\providecommand \EOS [0]{\spacefactor3000\relax}%
\providecommand \BibitemShut  [1]{\csname bibitem#1\endcsname}%
\let\auto@bib@innerbib\@empty
\bibitem [{\citenamefont {Weisskopf}\ and\ \citenamefont
  {Wigner}(1930)}]{WeisskopfWigner}%
  \BibitemOpen
  \bibfield  {author} {\bibinfo {author} {\bibfnamefont {V.}~\bibnamefont
  {Weisskopf}}\ and\ \bibinfo {author} {\bibfnamefont {E.}~\bibnamefont
  {Wigner}},\ }\href {\doibase 10.1007/BF01336768} {\bibfield  {journal}
  {\bibinfo  {journal} {Zeitschrift f{\"u}r Physik}\ }\textbf {\bibinfo
  {volume} {63}},\ \bibinfo {pages} {54} (\bibinfo {year} {1930})}\BibitemShut
  {NoStop}%
\bibitem [{\citenamefont {Breuer}\ and\ \citenamefont
  {Petruccione}(2007)}]{breuerPetruccione2007}%
  \BibitemOpen
  \bibfield  {author} {\bibinfo {author} {\bibfnamefont {H.~P.}\ \bibnamefont
  {Breuer}}\ and\ \bibinfo {author} {\bibfnamefont {F.}~\bibnamefont
  {Petruccione}},\ }\href {\doibase 10.1093/acprof:oso/9780199213900.001.0001}
  {\emph {\bibinfo {title} {The {{Theory}} of {{Open Quantum Systems}}}}},\
  \bibinfo {edition} {1st}\ ed.,\ Vol.\ \bibinfo {volume} {9780199213}\
  (\bibinfo  {publisher} {{OUP Oxford}},\ \bibinfo {year} {2007})\BibitemShut
  {NoStop}%
\bibitem [{\citenamefont {Raimond}\ \emph {et~al.}(2001)\citenamefont
  {Raimond}, \citenamefont {Brune},\ and\ \citenamefont
  {Haroche}}]{raimond2001manipulating}%
  \BibitemOpen
  \bibfield  {author} {\bibinfo {author} {\bibfnamefont {J.~M.}\ \bibnamefont
  {Raimond}}, \bibinfo {author} {\bibfnamefont {M.}~\bibnamefont {Brune}}, \
  and\ \bibinfo {author} {\bibfnamefont {S.}~\bibnamefont {Haroche}},\ }\href
  {\doibase 10.1103/RevModPhys.73.565} {\bibfield  {journal} {\bibinfo
  {journal} {Rev. Mod. Phys.}\ }\textbf {\bibinfo {volume} {73}},\ \bibinfo
  {pages} {565} (\bibinfo {year} {2001})}\BibitemShut {NoStop}%
\bibitem [{\citenamefont {Blais}\ \emph {et~al.}(2021)\citenamefont {Blais},
  \citenamefont {Grimsmo}, \citenamefont {Girvin},\ and\ \citenamefont
  {Wallraff}}]{blais2021circuit}%
  \BibitemOpen
  \bibfield  {author} {\bibinfo {author} {\bibfnamefont {A.}~\bibnamefont
  {Blais}}, \bibinfo {author} {\bibfnamefont {A.~L.}\ \bibnamefont {Grimsmo}},
  \bibinfo {author} {\bibfnamefont {S.~M.}\ \bibnamefont {Girvin}}, \ and\
  \bibinfo {author} {\bibfnamefont {A.}~\bibnamefont {Wallraff}},\ }\href
  {\doibase 10.1103/RevModPhys.93.025005} {\bibfield  {journal} {\bibinfo
  {journal} {Rev. Mod. Phys.}\ }\textbf {\bibinfo {volume} {93}},\ \bibinfo
  {pages} {025005} (\bibinfo {year} {2021})}\BibitemShut {NoStop}%
\bibitem [{\citenamefont {Liu}\ and\ \citenamefont
  {Houck}(2017)}]{Liu2017Quantum}%
  \BibitemOpen
  \bibfield  {author} {\bibinfo {author} {\bibfnamefont {Y.}~\bibnamefont
  {Liu}}\ and\ \bibinfo {author} {\bibfnamefont {A.~A.}\ \bibnamefont
  {Houck}},\ }\href {\doibase 10.1038/nphys3834} {\bibfield  {journal}
  {\bibinfo  {journal} {Nature Physics}\ }\textbf {\bibinfo {volume} {13}},\
  \bibinfo {pages} {48} (\bibinfo {year} {2017})}\BibitemShut {NoStop}%
\bibitem [{\citenamefont {Sheremet}\ \emph {et~al.}(2021)\citenamefont
  {Sheremet}, \citenamefont {Petrov}, \citenamefont {Iorsh}, \citenamefont
  {Poshakinskiy},\ and\ \citenamefont {Poddubny}}]{sheremet2021waveguide}%
  \BibitemOpen
  \bibfield  {author} {\bibinfo {author} {\bibfnamefont {A.~S.}\ \bibnamefont
  {Sheremet}}, \bibinfo {author} {\bibfnamefont {M.~I.}\ \bibnamefont
  {Petrov}}, \bibinfo {author} {\bibfnamefont {I.~V.}\ \bibnamefont {Iorsh}},
  \bibinfo {author} {\bibfnamefont {A.~V.}\ \bibnamefont {Poshakinskiy}}, \
  and\ \bibinfo {author} {\bibfnamefont {A.~N.}\ \bibnamefont {Poddubny}},\
  }\href {\doibase 10.48550/ARXIV.2103.06824} {\enquote {\bibinfo {title}
  {Waveguide quantum electrodynamics: collective radiance and photon-photon
  correlations},}\ } (\bibinfo {year} {2021})\BibitemShut {NoStop}%
\bibitem [{\citenamefont {Goban}\ \emph {et~al.}(2014)\citenamefont {Goban},
  \citenamefont {Hung}, \citenamefont {Yu}, \citenamefont {Hood}, \citenamefont
  {Muniz}, \citenamefont {Lee}, \citenamefont {Martin}, \citenamefont
  {McClung}, \citenamefont {Choi}, \citenamefont {Chang} \emph
  {et~al.}}]{goban2014atom}%
  \BibitemOpen
  \bibfield  {author} {\bibinfo {author} {\bibfnamefont {A.}~\bibnamefont
  {Goban}}, \bibinfo {author} {\bibfnamefont {C.-L.}\ \bibnamefont {Hung}},
  \bibinfo {author} {\bibfnamefont {S.-P.}\ \bibnamefont {Yu}}, \bibinfo
  {author} {\bibfnamefont {J.}~\bibnamefont {Hood}}, \bibinfo {author}
  {\bibfnamefont {J.}~\bibnamefont {Muniz}}, \bibinfo {author} {\bibfnamefont
  {J.}~\bibnamefont {Lee}}, \bibinfo {author} {\bibfnamefont {M.}~\bibnamefont
  {Martin}}, \bibinfo {author} {\bibfnamefont {A.}~\bibnamefont {McClung}},
  \bibinfo {author} {\bibfnamefont {K.}~\bibnamefont {Choi}}, \bibinfo {author}
  {\bibfnamefont {D.~E.}\ \bibnamefont {Chang}},  \emph {et~al.},\ }\href@noop
  {} {\bibfield  {journal} {\bibinfo  {journal} {Nature communications}\
  }\textbf {\bibinfo {volume} {5}},\ \bibinfo {pages} {1} (\bibinfo {year}
  {2014})}\BibitemShut {NoStop}%
\bibitem [{\citenamefont {Hood}\ \emph {et~al.}(2016)\citenamefont {Hood},
  \citenamefont {Goban}, \citenamefont {Asenjo-Garcia}, \citenamefont {Lu},
  \citenamefont {Yu}, \citenamefont {Chang},\ and\ \citenamefont
  {Kimble}}]{hood2016atom}%
  \BibitemOpen
  \bibfield  {author} {\bibinfo {author} {\bibfnamefont {J.~D.}\ \bibnamefont
  {Hood}}, \bibinfo {author} {\bibfnamefont {A.}~\bibnamefont {Goban}},
  \bibinfo {author} {\bibfnamefont {A.}~\bibnamefont {Asenjo-Garcia}}, \bibinfo
  {author} {\bibfnamefont {M.}~\bibnamefont {Lu}}, \bibinfo {author}
  {\bibfnamefont {S.-P.}\ \bibnamefont {Yu}}, \bibinfo {author} {\bibfnamefont
  {D.~E.}\ \bibnamefont {Chang}}, \ and\ \bibinfo {author} {\bibfnamefont
  {H.~J.}\ \bibnamefont {Kimble}},\ }\href {\doibase 10.1073/pnas.1603788113}
  {\bibfield  {journal} {\bibinfo  {journal} {Proceedings of the National
  Academy of Sciences}\ }\textbf {\bibinfo {volume} {113}},\ \bibinfo {pages}
  {10507} (\bibinfo {year} {2016})},\ \Eprint
  {http://arxiv.org/abs/https://www.pnas.org/doi/pdf/10.1073/pnas.1603788113}
  {https://www.pnas.org/doi/pdf/10.1073/pnas.1603788113} \BibitemShut {NoStop}%
\bibitem [{\citenamefont {Krinner}\ \emph {et~al.}(2018)\citenamefont
  {Krinner}, \citenamefont {Stewart}, \citenamefont {Pazmi{\~n}o},
  \citenamefont {Kwon},\ and\ \citenamefont
  {Schneble}}]{krinner2018spontaneous}%
  \BibitemOpen
  \bibfield  {author} {\bibinfo {author} {\bibfnamefont {L.}~\bibnamefont
  {Krinner}}, \bibinfo {author} {\bibfnamefont {M.}~\bibnamefont {Stewart}},
  \bibinfo {author} {\bibfnamefont {A.}~\bibnamefont {Pazmi{\~n}o}}, \bibinfo
  {author} {\bibfnamefont {J.}~\bibnamefont {Kwon}}, \ and\ \bibinfo {author}
  {\bibfnamefont {D.}~\bibnamefont {Schneble}},\ }\href {\doibase
  10.1038/s41586-018-0348-z} {\bibfield  {journal} {\bibinfo  {journal}
  {Nature}\ }\textbf {\bibinfo {volume} {559}},\ \bibinfo {pages} {589}
  (\bibinfo {year} {2018})}\BibitemShut {NoStop}%
\bibitem [{\citenamefont {Stewart}\ \emph {et~al.}(2020)\citenamefont
  {Stewart}, \citenamefont {Kwon}, \citenamefont {Lanuza},\ and\ \citenamefont
  {Schneble}}]{stewart2020dynamics}%
  \BibitemOpen
  \bibfield  {author} {\bibinfo {author} {\bibfnamefont {M.}~\bibnamefont
  {Stewart}}, \bibinfo {author} {\bibfnamefont {J.}~\bibnamefont {Kwon}},
  \bibinfo {author} {\bibfnamefont {A.}~\bibnamefont {Lanuza}}, \ and\ \bibinfo
  {author} {\bibfnamefont {D.}~\bibnamefont {Schneble}},\ }\href {\doibase
  10.1103/PhysRevResearch.2.043307} {\bibfield  {journal} {\bibinfo  {journal}
  {Phys. Rev. Research}\ }\textbf {\bibinfo {volume} {2}},\ \bibinfo {pages}
  {043307} (\bibinfo {year} {2020})}\BibitemShut {NoStop}%
\bibitem [{\citenamefont {John}\ and\ \citenamefont
  {Wang}(1990)}]{john1990quantum}%
  \BibitemOpen
  \bibfield  {author} {\bibinfo {author} {\bibfnamefont {S.}~\bibnamefont
  {John}}\ and\ \bibinfo {author} {\bibfnamefont {J.}~\bibnamefont {Wang}},\
  }\href {\doibase 10.1103/PhysRevLett.64.2418} {\bibfield  {journal} {\bibinfo
   {journal} {Phys. Rev. Lett.}\ }\textbf {\bibinfo {volume} {64}},\ \bibinfo
  {pages} {2418} (\bibinfo {year} {1990})}\BibitemShut {NoStop}%
\bibitem [{\citenamefont {John}\ and\ \citenamefont
  {Quang}(1994)}]{john1994spontaneous}%
  \BibitemOpen
  \bibfield  {author} {\bibinfo {author} {\bibfnamefont {S.}~\bibnamefont
  {John}}\ and\ \bibinfo {author} {\bibfnamefont {T.}~\bibnamefont {Quang}},\
  }\href {\doibase 10.1103/PhysRevA.50.1764} {\bibfield  {journal} {\bibinfo
  {journal} {Phys. Rev. A}\ }\textbf {\bibinfo {volume} {50}},\ \bibinfo
  {pages} {1764} (\bibinfo {year} {1994})}\BibitemShut {NoStop}%
\bibitem [{\citenamefont {Mogilevtsev}\ \emph {et~al.}(2005)\citenamefont
  {Mogilevtsev}, \citenamefont {Kilin}, \citenamefont {Cavalcanti},\ and\
  \citenamefont {Hickmann}}]{mogilevtsev2005inreservoir}%
  \BibitemOpen
  \bibfield  {author} {\bibinfo {author} {\bibfnamefont {D.}~\bibnamefont
  {Mogilevtsev}}, \bibinfo {author} {\bibfnamefont {S.}~\bibnamefont {Kilin}},
  \bibinfo {author} {\bibfnamefont {S.}~\bibnamefont {Cavalcanti}}, \ and\
  \bibinfo {author} {\bibfnamefont {J.}~\bibnamefont {Hickmann}},\ }\href
  {\doibase 10.1103/PhysRevA.72.043817} {\bibfield  {journal} {\bibinfo
  {journal} {Phys. Rev. A}\ }\textbf {\bibinfo {volume} {72}} (\bibinfo {year}
  {2005}),\ 10.1103/PhysRevA.72.043817}\BibitemShut {NoStop}%
\bibitem [{\citenamefont {Mogilevtsev}\ and\ \citenamefont
  {Kilin}(2008)}]{mogilevtsev2008effective}%
  \BibitemOpen
  \bibfield  {author} {\bibinfo {author} {\bibfnamefont {D.}~\bibnamefont
  {Mogilevtsev}}\ and\ \bibinfo {author} {\bibfnamefont {S.}~\bibnamefont
  {Kilin}},\ }\href {\doibase 10.1103/PhysRevA.78.033808} {\bibfield  {journal}
  {\bibinfo  {journal} {Phys. Rev. A}\ }\textbf {\bibinfo {volume} {78}},\
  \bibinfo {pages} {033808} (\bibinfo {year} {2008})}\BibitemShut {NoStop}%
\bibitem [{\citenamefont {Lambropoulos}\ \emph {et~al.}(2000)\citenamefont
  {Lambropoulos}, \citenamefont {Nikolopoulos}, \citenamefont {Nielsen},\ and\
  \citenamefont {Bay}}]{Lambropoulos_2000}%
  \BibitemOpen
  \bibfield  {author} {\bibinfo {author} {\bibfnamefont {P.}~\bibnamefont
  {Lambropoulos}}, \bibinfo {author} {\bibfnamefont {G.~M.}\ \bibnamefont
  {Nikolopoulos}}, \bibinfo {author} {\bibfnamefont {T.~R.}\ \bibnamefont
  {Nielsen}}, \ and\ \bibinfo {author} {\bibfnamefont {S.}~\bibnamefont
  {Bay}},\ }\href {\doibase 10.1088/0034-4885/63/4/201} {\bibfield  {journal}
  {\bibinfo  {journal} {Reports on Progress in Physics}\ }\textbf {\bibinfo
  {volume} {63}},\ \bibinfo {pages} {455} (\bibinfo {year} {2000})}\BibitemShut
  {NoStop}%
\bibitem [{\citenamefont {Shi}\ \emph {et~al.}(2016)\citenamefont {Shi},
  \citenamefont {Wu}, \citenamefont {Gonz\'alez-Tudela},\ and\ \citenamefont
  {Cirac}}]{shi2016bound}%
  \BibitemOpen
  \bibfield  {author} {\bibinfo {author} {\bibfnamefont {T.}~\bibnamefont
  {Shi}}, \bibinfo {author} {\bibfnamefont {Y.-H.}\ \bibnamefont {Wu}},
  \bibinfo {author} {\bibfnamefont {A.}~\bibnamefont {Gonz\'alez-Tudela}}, \
  and\ \bibinfo {author} {\bibfnamefont {J.~I.}\ \bibnamefont {Cirac}},\ }\href
  {\doibase 10.1103/PhysRevX.6.021027} {\bibfield  {journal} {\bibinfo
  {journal} {Phys. Rev. X}\ }\textbf {\bibinfo {volume} {6}},\ \bibinfo {pages}
  {021027} (\bibinfo {year} {2016})}\BibitemShut {NoStop}%
\bibitem [{\citenamefont {{Gonz{\'a}lez-Tudela}}\ and\ \citenamefont
  {Cirac}(2017{\natexlab{a}})}]{gonzalez-tudelaCirac2017}%
  \BibitemOpen
  \bibfield  {author} {\bibinfo {author} {\bibfnamefont {A.}~\bibnamefont
  {{Gonz{\'a}lez-Tudela}}}\ and\ \bibinfo {author} {\bibfnamefont {J.~I.}\
  \bibnamefont {Cirac}},\ }\href {\doibase 10.1103/PhysRevLett.119.143602}
  {\bibfield  {journal} {\bibinfo  {journal} {Phys. Rev. Lett.}\ }\textbf
  {\bibinfo {volume} {119}},\ \bibinfo {pages} {143602} (\bibinfo {year}
  {2017}{\natexlab{a}})}\BibitemShut {NoStop}%
\bibitem [{\citenamefont {{Gonz{\'a}lez-Tudela}}\ and\ \citenamefont
  {Cirac}(2017{\natexlab{b}})}]{gonzalez-tudelaCirac2017a}%
  \BibitemOpen
  \bibfield  {author} {\bibinfo {author} {\bibfnamefont {A.}~\bibnamefont
  {{Gonz{\'a}lez-Tudela}}}\ and\ \bibinfo {author} {\bibfnamefont {J.~I.}\
  \bibnamefont {Cirac}},\ }\href {\doibase 10.1103/PhysRevA.96.043811}
  {\bibfield  {journal} {\bibinfo  {journal} {Phys. Rev. A}\ }\textbf {\bibinfo
  {volume} {96}},\ \bibinfo {pages} {043811} (\bibinfo {year}
  {2017}{\natexlab{b}})}\BibitemShut {NoStop}%
\bibitem [{\citenamefont {Frisk~Kockum}\ \emph {et~al.}(2019)\citenamefont
  {Frisk~Kockum}, \citenamefont {Miranowicz}, \citenamefont {De~Liberato},
  \citenamefont {Savasta},\ and\ \citenamefont {Nori}}]{kockum2019ultra}%
  \BibitemOpen
  \bibfield  {author} {\bibinfo {author} {\bibfnamefont {A.}~\bibnamefont
  {Frisk~Kockum}}, \bibinfo {author} {\bibfnamefont {A.}~\bibnamefont
  {Miranowicz}}, \bibinfo {author} {\bibfnamefont {S.}~\bibnamefont
  {De~Liberato}}, \bibinfo {author} {\bibfnamefont {S.}~\bibnamefont
  {Savasta}}, \ and\ \bibinfo {author} {\bibfnamefont {F.}~\bibnamefont
  {Nori}},\ }\href {\doibase 10.1038/s42254-018-0006-2} {\bibfield  {journal}
  {\bibinfo  {journal} {Nature Reviews Physics}\ }\textbf {\bibinfo {volume}
  {1}},\ \bibinfo {pages} {19} (\bibinfo {year} {2019})}\BibitemShut {NoStop}%
\bibitem [{\citenamefont {Forn-D\'{\i}az}\ \emph {et~al.}(2019)\citenamefont
  {Forn-D\'{\i}az}, \citenamefont {Lamata}, \citenamefont {Rico}, \citenamefont
  {Kono},\ and\ \citenamefont {Solano}}]{forndiaz2019ultra}%
  \BibitemOpen
  \bibfield  {author} {\bibinfo {author} {\bibfnamefont {P.}~\bibnamefont
  {Forn-D\'{\i}az}}, \bibinfo {author} {\bibfnamefont {L.}~\bibnamefont
  {Lamata}}, \bibinfo {author} {\bibfnamefont {E.}~\bibnamefont {Rico}},
  \bibinfo {author} {\bibfnamefont {J.}~\bibnamefont {Kono}}, \ and\ \bibinfo
  {author} {\bibfnamefont {E.}~\bibnamefont {Solano}},\ }\href {\doibase
  10.1103/RevModPhys.91.025005} {\bibfield  {journal} {\bibinfo  {journal}
  {Rev. Mod. Phys.}\ }\textbf {\bibinfo {volume} {91}},\ \bibinfo {pages}
  {025005} (\bibinfo {year} {2019})}\BibitemShut {NoStop}%
\bibitem [{\citenamefont {Sanchez-Burillo}\ \emph {et~al.}(2014)\citenamefont
  {Sanchez-Burillo}, \citenamefont {Zueco}, \citenamefont {Garcia-Ripoll},\
  and\ \citenamefont {Martin-Moreno}}]{Sanchez_Burillo_2014}%
  \BibitemOpen
  \bibfield  {author} {\bibinfo {author} {\bibfnamefont {E.}~\bibnamefont
  {Sanchez-Burillo}}, \bibinfo {author} {\bibfnamefont {D.}~\bibnamefont
  {Zueco}}, \bibinfo {author} {\bibfnamefont {J.~J.}\ \bibnamefont
  {Garcia-Ripoll}}, \ and\ \bibinfo {author} {\bibfnamefont {L.}~\bibnamefont
  {Martin-Moreno}},\ }\href {\doibase 10.1103/physrevlett.113.263604}
  {\bibfield  {journal} {\bibinfo  {journal} {Physical Review Letters}\
  }\textbf {\bibinfo {volume} {113}} (\bibinfo {year} {2014}),\
  10.1103/physrevlett.113.263604}\BibitemShut {NoStop}%
\bibitem [{\citenamefont {D\'{\i}az-Camacho}\ \emph {et~al.}(2016)\citenamefont
  {D\'{\i}az-Camacho}, \citenamefont {Bermudez},\ and\ \citenamefont
  {Garc\'{\i}a-Ripoll}}]{diazcamacho2016dynamical}%
  \BibitemOpen
  \bibfield  {author} {\bibinfo {author} {\bibfnamefont {G.}~\bibnamefont
  {D\'{\i}az-Camacho}}, \bibinfo {author} {\bibfnamefont {A.}~\bibnamefont
  {Bermudez}}, \ and\ \bibinfo {author} {\bibfnamefont {J.~J.}\ \bibnamefont
  {Garc\'{\i}a-Ripoll}},\ }\href {\doibase 10.1103/PhysRevA.93.043843}
  {\bibfield  {journal} {\bibinfo  {journal} {Phys. Rev. A}\ }\textbf {\bibinfo
  {volume} {93}},\ \bibinfo {pages} {043843} (\bibinfo {year}
  {2016})}\BibitemShut {NoStop}%
\bibitem [{\citenamefont {Rom{\'{a}}n-Roche}\ \emph {et~al.}(2020)\citenamefont
  {Rom{\'{a}}n-Roche}, \citenamefont {S{\'{a}}nchez-Burillo},\ and\
  \citenamefont {Zueco}}]{Rom_n_Roche_2020}%
  \BibitemOpen
  \bibfield  {author} {\bibinfo {author} {\bibfnamefont {J.}~\bibnamefont
  {Rom{\'{a}}n-Roche}}, \bibinfo {author} {\bibfnamefont {E.}~\bibnamefont
  {S{\'{a}}nchez-Burillo}}, \ and\ \bibinfo {author} {\bibfnamefont
  {D.}~\bibnamefont {Zueco}},\ }\href {\doibase 10.1103/physreva.102.023702}
  {\bibfield  {journal} {\bibinfo  {journal} {Physical Review A}\ }\textbf
  {\bibinfo {volume} {102}} (\bibinfo {year} {2020}),\
  10.1103/physreva.102.023702}\BibitemShut {NoStop}%
\bibitem [{\citenamefont {Ashida}\ \emph {et~al.}(2022)\citenamefont {Ashida},
  \citenamefont {Yokota}, \citenamefont {\ifmmode \dot{I}\else
  \.{I}\fi{}mamo\ifmmode~\breve{g}\else \u{g}\fi{}lu},\ and\ \citenamefont
  {Demler}}]{Ashida2022Nonperturbative}%
  \BibitemOpen
  \bibfield  {author} {\bibinfo {author} {\bibfnamefont {Y.}~\bibnamefont
  {Ashida}}, \bibinfo {author} {\bibfnamefont {T.}~\bibnamefont {Yokota}},
  \bibinfo {author} {\bibfnamefont {A.~m.~c.}\ \bibnamefont {\ifmmode
  \dot{I}\else \.{I}\fi{}mamo\ifmmode~\breve{g}\else \u{g}\fi{}lu}}, \ and\
  \bibinfo {author} {\bibfnamefont {E.}~\bibnamefont {Demler}},\ }\href
  {\doibase 10.1103/PhysRevResearch.4.023194} {\bibfield  {journal} {\bibinfo
  {journal} {Phys. Rev. Research}\ }\textbf {\bibinfo {volume} {4}},\ \bibinfo
  {pages} {023194} (\bibinfo {year} {2022})}\BibitemShut {NoStop}%
\bibitem [{\citenamefont {Terradas-Briansó}\ \emph {et~al.}(2022)\citenamefont
  {Terradas-Briansó}, \citenamefont {González-Gutiérrez}, \citenamefont
  {Nori}, \citenamefont {Martín-Moreno},\ and\ \citenamefont
  {Zueco}}]{terradas2022ultrastrong}%
  \BibitemOpen
  \bibfield  {author} {\bibinfo {author} {\bibfnamefont {S.}~\bibnamefont
  {Terradas-Briansó}}, \bibinfo {author} {\bibfnamefont {C.~A.}\ \bibnamefont
  {González-Gutiérrez}}, \bibinfo {author} {\bibfnamefont {F.}~\bibnamefont
  {Nori}}, \bibinfo {author} {\bibfnamefont {L.}~\bibnamefont
  {Martín-Moreno}}, \ and\ \bibinfo {author} {\bibfnamefont {D.}~\bibnamefont
  {Zueco}},\ }\href {\doibase 10.48550/ARXIV.2205.07915} {\enquote {\bibinfo
  {title} {Ultrastrong waveguide qed with giant atoms},}\ } (\bibinfo {year}
  {2022})\BibitemShut {NoStop}%
\bibitem [{\citenamefont {Scarlatella}\ and\ \citenamefont
  {Schiro}(2021)}]{scarlatella2021noncrossing}%
  \BibitemOpen
  \bibfield  {author} {\bibinfo {author} {\bibfnamefont {O.}~\bibnamefont
  {Scarlatella}}\ and\ \bibinfo {author} {\bibfnamefont {M.}~\bibnamefont
  {Schiro}},\ }\href {\doibase 10.48550/ARXIV.2107.05553} {\enquote {\bibinfo
  {title} {Non-crossing dynamical maps for open quantum systems},}\ } (\bibinfo
  {year} {2021})\BibitemShut {NoStop}%
\bibitem [{\citenamefont {Gonz{\'{a}}lez-Tudela}\ and\ \citenamefont
  {Cirac}(2017{\natexlab{a}})}]{Gonz_lez_Tudela_2017_short}%
  \BibitemOpen
  \bibfield  {author} {\bibinfo {author} {\bibfnamefont {A.}~\bibnamefont
  {Gonz{\'{a}}lez-Tudela}}\ and\ \bibinfo {author} {\bibfnamefont
  {J.}~\bibnamefont {Cirac}},\ }\href {\doibase 10.1103/physrevlett.119.143602}
  {\bibfield  {journal} {\bibinfo  {journal} {Physical Review Letters}\
  }\textbf {\bibinfo {volume} {119}} (\bibinfo {year} {2017}{\natexlab{a}}),\
  10.1103/physrevlett.119.143602}\BibitemShut {NoStop}%
\bibitem [{\citenamefont {Gonz{\'{a}}lez-Tudela}\ and\ \citenamefont
  {Cirac}(2017{\natexlab{b}})}]{Gonz_lez_Tudela_2017}%
  \BibitemOpen
  \bibfield  {author} {\bibinfo {author} {\bibfnamefont {A.}~\bibnamefont
  {Gonz{\'{a}}lez-Tudela}}\ and\ \bibinfo {author} {\bibfnamefont {J.~I.}\
  \bibnamefont {Cirac}},\ }\href {\doibase 10.1103/physreva.96.043811}
  {\bibfield  {journal} {\bibinfo  {journal} {Physical Review A}\ }\textbf
  {\bibinfo {volume} {96}} (\bibinfo {year} {2017}{\natexlab{b}}),\
  10.1103/physreva.96.043811}\BibitemShut {NoStop}%
\bibitem [{\citenamefont {Gonz{\'{a}}lez-Tudela}\ and\ \citenamefont
  {Cirac}(2018)}]{Gonz_lez_Tudela_2018}%
  \BibitemOpen
  \bibfield  {author} {\bibinfo {author} {\bibfnamefont {A.}~\bibnamefont
  {Gonz{\'{a}}lez-Tudela}}\ and\ \bibinfo {author} {\bibfnamefont {J.~I.}\
  \bibnamefont {Cirac}},\ }\href {\doibase 10.1103/physreva.97.043831}
  {\bibfield  {journal} {\bibinfo  {journal} {Physical Review A}\ }\textbf
  {\bibinfo {volume} {97}} (\bibinfo {year} {2018}),\
  10.1103/physreva.97.043831}\BibitemShut {NoStop}%
\bibitem [{\citenamefont {Yuan}\ \emph {et~al.}(2017)\citenamefont {Yuan},
  \citenamefont {Xing}, \citenamefont {Kuang},\ and\ \citenamefont
  {Yi}}]{yuanYi2017}%
  \BibitemOpen
  \bibfield  {author} {\bibinfo {author} {\bibfnamefont {J.-B.}\ \bibnamefont
  {Yuan}}, \bibinfo {author} {\bibfnamefont {H.-J.}\ \bibnamefont {Xing}},
  \bibinfo {author} {\bibfnamefont {L.-M.}\ \bibnamefont {Kuang}}, \ and\
  \bibinfo {author} {\bibfnamefont {S.}~\bibnamefont {Yi}},\ }\href {\doibase
  10.1103/PhysRevA.95.033610} {\bibfield  {journal} {\bibinfo  {journal} {Phys.
  Rev. A}\ }\textbf {\bibinfo {volume} {95}},\ \bibinfo {pages} {033610}
  (\bibinfo {year} {2017})}\BibitemShut {NoStop}%
\bibitem [{\citenamefont {D{\'{\i}}az-Camacho}\ \emph
  {et~al.}(2016)\citenamefont {D{\'{\i}}az-Camacho}, \citenamefont {Bermudez},\
  and\ \citenamefont {Garc{\'{\i}}a-Ripoll}}]{D_az_Camacho_2016}%
  \BibitemOpen
  \bibfield  {author} {\bibinfo {author} {\bibfnamefont {G.}~\bibnamefont
  {D{\'{\i}}az-Camacho}}, \bibinfo {author} {\bibfnamefont {A.}~\bibnamefont
  {Bermudez}}, \ and\ \bibinfo {author} {\bibfnamefont {J.~J.}\ \bibnamefont
  {Garc{\'{\i}}a-Ripoll}},\ }\href {\doibase 10.1103/physreva.93.043843}
  {\bibfield  {journal} {\bibinfo  {journal} {Physical Review A}\ }\textbf
  {\bibinfo {volume} {93}} (\bibinfo {year} {2016}),\
  10.1103/physreva.93.043843}\BibitemShut {NoStop}%
\bibitem [{\citenamefont {Kopp}\ and\ \citenamefont
  {Hur}(2007)}]{kopp2007universal}%
  \BibitemOpen
  \bibfield  {author} {\bibinfo {author} {\bibfnamefont {A.}~\bibnamefont
  {Kopp}}\ and\ \bibinfo {author} {\bibfnamefont {K.~L.}\ \bibnamefont {Hur}},\
  }\href {\doibase 10.1103/PhysRevLett.98.220401} {\bibfield  {journal}
  {\bibinfo  {journal} {Phys. Rev. Lett.}\ }\textbf {\bibinfo {volume} {98}},\
  \bibinfo {pages} {220401} (\bibinfo {year} {2007})}\BibitemShut {NoStop}%
\bibitem [{\citenamefont {Le~Hur}\ \emph {et~al.}(2007)\citenamefont {Le~Hur},
  \citenamefont {Doucet-Beaupr\'e},\ and\ \citenamefont
  {Hofstetter}}]{lehur2007entanglement}%
  \BibitemOpen
  \bibfield  {author} {\bibinfo {author} {\bibfnamefont {K.}~\bibnamefont
  {Le~Hur}}, \bibinfo {author} {\bibfnamefont {P.}~\bibnamefont
  {Doucet-Beaupr\'e}}, \ and\ \bibinfo {author} {\bibfnamefont
  {W.}~\bibnamefont {Hofstetter}},\ }\href {\doibase
  10.1103/PhysRevLett.99.126801} {\bibfield  {journal} {\bibinfo  {journal}
  {Phys. Rev. Lett.}\ }\textbf {\bibinfo {volume} {99}},\ \bibinfo {pages}
  {126801} (\bibinfo {year} {2007})}\BibitemShut {NoStop}%
\bibitem [{\citenamefont {Hur}(2008)}]{HUR20082208}%
  \BibitemOpen
  \bibfield  {author} {\bibinfo {author} {\bibfnamefont {K.~L.}\ \bibnamefont
  {Hur}},\ }\href {\doibase https://doi.org/10.1016/j.aop.2007.12.003}
  {\bibfield  {journal} {\bibinfo  {journal} {Annals of Physics}\ }\textbf
  {\bibinfo {volume} {323}},\ \bibinfo {pages} {2208} (\bibinfo {year}
  {2008})}\BibitemShut {NoStop}%
\bibitem [{\citenamefont {Kilin}\ and\ \citenamefont
  {Mogilevtsev}(1992)}]{kilin1992freezing}%
  \BibitemOpen
  \bibfield  {author} {\bibinfo {author} {\bibfnamefont {S.}~\bibnamefont
  {Kilin}}\ and\ \bibinfo {author} {\bibfnamefont {D.}~\bibnamefont
  {Mogilevtsev}},\ }\href@noop {} {\bibfield  {journal} {\bibinfo  {journal}
  {Laser Phys.}\ }\textbf {\bibinfo {volume} {2}},\ \bibinfo {pages} {153}
  (\bibinfo {year} {1992})}\BibitemShut {NoStop}%
\bibitem [{\citenamefont {{Kilin}}\ and\ \citenamefont
  {{Mogilevtsev}}(1993)}]{1993OptSp74579K}%
  \BibitemOpen
  \bibfield  {author} {\bibinfo {author} {\bibfnamefont {S.~Y.}\ \bibnamefont
  {{Kilin}}}\ and\ \bibinfo {author} {\bibfnamefont {D.~S.}\ \bibnamefont
  {{Mogilevtsev}}},\ }\href@noop {} {\bibfield  {journal} {\bibinfo  {journal}
  {Optics and Spectroscopy}\ }\textbf {\bibinfo {volume} {74}},\ \bibinfo
  {pages} {579} (\bibinfo {year} {1993})}\BibitemShut {NoStop}%
\bibitem [{\citenamefont {John}\ and\ \citenamefont
  {Quang}(1995)}]{john1995localization}%
  \BibitemOpen
  \bibfield  {author} {\bibinfo {author} {\bibfnamefont {S.}~\bibnamefont
  {John}}\ and\ \bibinfo {author} {\bibfnamefont {T.}~\bibnamefont {Quang}},\
  }\href {\doibase 10.1103/PhysRevLett.74.3419} {\bibfield  {journal} {\bibinfo
   {journal} {Phys. Rev. Lett.}\ }\textbf {\bibinfo {volume} {74}},\ \bibinfo
  {pages} {3419} (\bibinfo {year} {1995})}\BibitemShut {NoStop}%
\end{thebibliography}

\newpage
\onecolumngrid
\appendix

\section{Dyson equation for RWA dynamics}\label{app:dyson}
In the same spirit as in the main text, we can write the dyson equation and more particularly the self energy for the Hamiltonian $\m{H}_{RWA}$:
\begin{align}
    \Sigma_{NCA}^{RWA}(t_1,t_2) = &- \Gamma_{++}(t_2 - t_1) X_{+} \m{V}(t_1,t_2) X_+^\dagger \notag \\ & - \Gamma_{--}(t_1 - t_2) X_{-}^\dagger \m{V}(t_1,t_2) X_- \notag \\ &  +\Gamma_{+-}(t_1 - t_2) X_{+}^\dagger \m{V}(t_1,t_2) X_- \notag \\ & + \Gamma_{+-}(t_2 - t_1) X_{-} \m{V}(t_1,t_2) X_+^\dagger
\end{align}
where the operator X is $X=\sigma^+ = \sigma^x + i\sigma^y$ and $X_\gamma$ are superoperators labeled by indices $\gamma\in\{+,-\}$ such that $X_+ = X \bullet$ and $X_- =  \bullet X$.
The bath correlation functions are given by
\begin{align}
    \Gamma_{\gamma \gamma'}(t-t') =g^2 \sum_{\bb{k}} Tr_B \left(\m{T}_c a_{\bb{k},\gamma}^\dagger(t) a_{\bb{k},\gamma'}(t') \rho_{B}(0) \right)
\end{align}

\section{Validity of the rotating wave approximation}\label{app:validity}

In this section, we will determine the validity criterion for the RWA described in the main text.  For this, we will consider the evolution of the density matrix described by the following Hamiltonian:
\begin{align}
    \m{H} = \m{H}_{RW}+\m{H}_{CR} =\m{H}_{RWA} + g\sum_\bb{k} \sigma^+ a_\bb{k}^\dagger + \sigma^- a_\bb{k} 
\end{align}

we recall that the evolution of the density matrix is given by $\rho_{tot}(t) = e^{-i\m{H}t}\rho_{tot}(0)e^{i\m{H}t}$, where $e^{i\m{H}t}$ is the time evolution operator for the total Hamiltonian. By moving to the interaction picture according to the unperturbed Hamiltonian $\m{H}_{RWA}$, the following identites can be found
\begin{equation}
    \rho_{tot}(t) = e^{-i\m{H}t}\rho_{tot}(0)e^{i\m{H}t} = e^{-i\m{H}_{RWA}t}\m{T}_t e^{-i \int_0^t  \m{H}_{CR}(\tau) d\tau }  
    \rho_{tot}(0) 
    \tilde{\m{T}}_t e^{i \int_0^t  \m{H}_{CR}(\tau) d\tau } e^{i\m{H}_{RWA}t}
\end{equation}
here $\m{H}_{CR}(\tau) = e^{i\m{H}_{RWA}\tau}\m{H}_{CR}e^{-i\m{H}_{RWA}\tau}$ and  $\m{T}_t$ is the real-time ordering operator, which for any operator product changes the order such that each operator has only later operators to the left and earlier operators to the right. From this expression, we see that the rotating wave approximation is valid when the initial density matrix remains unchanged by the evolution according to the Hamiltonian $\m{H}_{CR}(t)$, i.e when:
\begin{align}
    \rho_{tot}^{CR}(t) = e^{-i \int_0^t  \m{H}_{CR}(\tau) d\tau }  \rho_{tot}(0)  e^{i \int_0^t  \m{H}_{CR}(\tau) d\tau } \approx \rho_{tot}(0)
\end{align}
Since this condition is in general not satisfied, we perform an expansion according to the orders in g, and we seek the condition such that the lowest orders are negligible. In order to manage this expansion, it is convenient to rewrite the evolution operator
\begin{align}
    \int_{0}^{t} \m{H}_{CR}(\tau) d\tau = \int_0^t \m{T}_t e^{i\int_0^t d\tau \m{H}_{SB}(t')dt'}e^{i\left( \m{H}_S + \m{H}_B\right)\tau}\m{H}_{CR} e^{-i\left( \m{H}_S + \m{H}_B\right)\tau} \tilde{\m{T}}_t e^{-i\int_0^t d\tau \m{H}_{SB}(t')dt'}
\end{align}
where we used the interaction picture on the real-time evolution operator of the Hamiltonian $\m{H}_{RWA} = \m{H}_S + \m{H}_B + \m{H}_{SB}$. By using the definitions of $\m{H}_B$ and $\m{H}_S$  from the main text, the following identities can be found for the part without coupling with the bath
\begin{align}
    e^{i\left( \m{H}_S + \m{H}_B\right)\tau}\m{H}_{CR} e^{-i\left( \m{H}_S + \m{H}_B\right)\tau} = g\sum_\bb{k} \sigma^+ a_{\bb{k}}^\dagger e^{i\left( \omega_q+ \Tilde{\omega}_\bb{k}\right)\tau} + \sigma^- a_\bb{k} e^{-i\left(\omega_q + \Tilde{\omega}_\bb{k} \right)\tau}
\end{align}
this expression simply corresponds to the rotating frame of the counter rotating terms. Moreover, since the couplings are linear in g, we can expand the exponentials for each order. For the first order we get
\begin{align}
    \rho_{tot}^{CR, (1)}(t) = ig \int_0^t d\tau \left[e^{i\left( \m{H}_S + \m{H}_B\right)\tau}\m{H}_{CR} e^{-i\left( \m{H}_S + \m{H}_B\right)\tau} \right] \rho_{tot}(0) -ig \rho_{tot}(0)\int_0^t d\tau \left[  e^{i\left( \m{H}_S + \m{H}_B\right)\tau}\m{H}_{CR} e^{-i\left( \m{H}_S + \m{H}_B\right)\tau} \right]
\end{align}
the order 1 in g does not contribute because initially we have the density matrix $\rho_{tot}(0) = \ket{\uparrow}\otimes \ket{\bb{0}}$. In the same way for the second order, we have 5 possible contributions for the dynamics of the Qubit:
\begin{align}
    &1)  -\frac{1}{2} \m{T}_t \int_{0}^t \int_{0}^t d\tau_1 d\tau_2 \m{H}_{CR}^{(1)}(\tau_1) \m{H}_{CR}^{(1)}(\tau_2) \rho_{tot}(0) \notag \\&2)  -\frac{1}{2} \rho_{tot}(0) \tilde{\m{T}}_t \int_{0}^t \int_{0}^t d\tau_1 d\tau_2 \m{H}_{CR}^{(1)}(\tau_1) \m{H}_{CR}^{(1)}(\tau_2) \notag \\&3) +\m{T}_t \int_0^t d\tau \m{H}_{CR}^{(1)}(\tau) \rho_{tot}(0) \tilde{\m{T}}_t \int_0^t d\tau \m{H}_{CR}^{(1)}(\tau) \notag \\ &4) -i\m{T}_t \int_0^t d\tau \m{H}_{CR}^{(2)}(\tau) \rho_{tot}(0) \notag \\&5) -i\rho_{tot}(0) \tilde{T}_t \int_0^t d\tau \m{H}_{CR}^{(2)}(\tau)
\end{align}
where in its expressions, we denote $\m{H}_{CR}^{(n)}=O\left(g^n \right)$,that is, we perform an expansion of the counter rotating term into the coupling constant g. Due to the initial state corresponding to a single excitation for the Qubit, one can easily convince oneself that the contributions (1-3) do not participate in the dynamics, because we have $\m{H}_{CR}^{(1)} =  e^{i\left( \m{H}_S + \m{H}_B\right)\tau}\m{H}_{CR} e^{-i\left( \m{H}_S + \m{H}_B\right)\tau}$.With regard to contribution 4) (in the same way for the contribution 5) we can rewrite it in the following form
\begin{align}
    -i \int_0^t d\tau \m{H}_{CR}^{(2)}(\tau) \rho_{tot}(0) = g^2 \sum_{\bb{k}_1,\bb{k}_2} \left[ \frac{e^{it \left(\tilde{\omega}_{\bb{k}_1} + \tilde{\omega}_{\bb{k}_2} \right)}-1}{\left( \tilde{\omega}_{\bb{k}_1} + \tilde{\omega}_{\bb{k}_2}\right) \left(\tilde{\omega}_{\bb{k}_2} - \omega_q \right)}  - \frac{e^{it \left(\tilde{\omega}_{\bb{k}_1} + \tilde{\omega}_{q} \right)}-1}{\left( \tilde{\omega}_{\bb{k}_1} + {\omega}_q\right) \left(\tilde{\omega}_{\bb{k}_2} - \omega_q \right)} \right] a_{\bb{k}_1}^\dagger a_{\bb{k}_2}^\dagger \sigma^+ \sigma^- \rho_{tot}(0)
\end{align}

So, the Rotating wave approximation is then valid when the following criterion is met:
\begin{align}
    g^2 \underset{\tilde{\omega}_{\bb{k}_1},\tilde{\omega}_{\bb{k}_2}}{max}{\left[\frac{1}{\left(\tilde{\omega}_{\bb{k}_1}+\tilde{\omega}_{\bb{k}_2}\right) \left(\omega_q + \tilde{\omega}_{\bb{k}_1} \right)}  \right]} \ll 1
\end{align}
using the definition of detuning $\Delta= \omega_q -\omega_*$ and remembering that we choose $\omega_*$ such that $\tilde{\omega}_\bb{k}>0$, for all $\bb{k}$, the expression can be reduced to 
\begin{align}
    g^2 \ll \left( 2\omega_* - 8J\right) \left(\Delta +2\omega_* - 4J \right)
\end{align}

\section{Comparison with Born Master equation}

\begin{figure}[!t]\label{fig:born}
    \centering
        \includegraphics[width=0.45\textwidth]{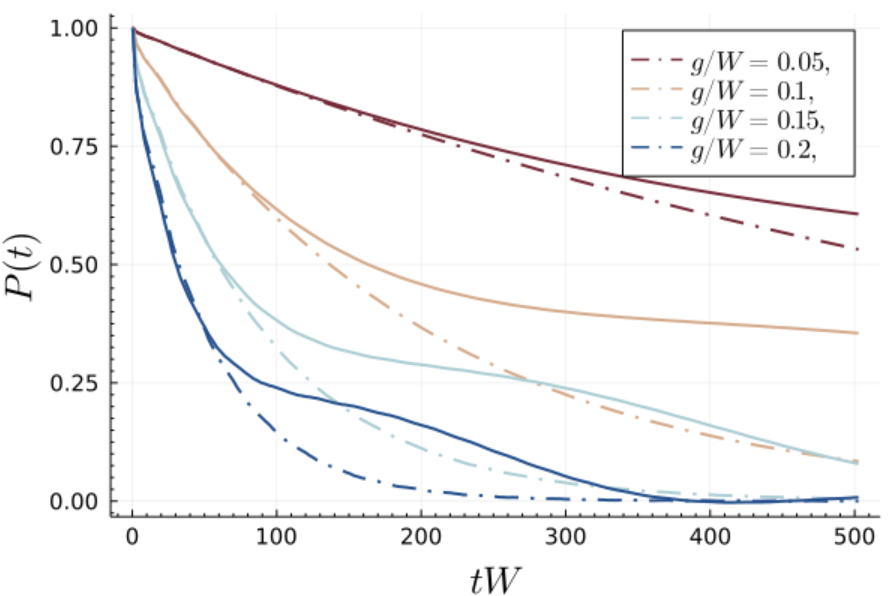}
        \includegraphics[width=0.45\textwidth]{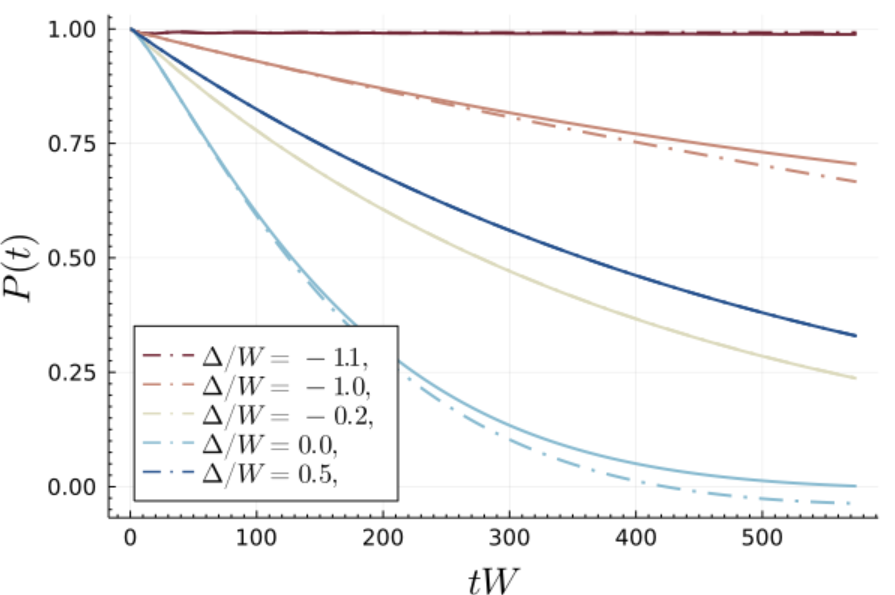}
    \caption{Comparison between the RWA dynamics obtained by NCA (solid line) and by the Born Master equation (dash line). The left panel corresponds to the edge of the band ($\Delta/W=-1$) for different values of the coupling g. As for the right panel it corespond has a fixed coupling $g=0.1J$ and different detuning $\Delta$. }
\end{figure}

In this appendix we show that many features of the emitter dynamics discussed in the main text are missed if one solves the system-bath dynamics within the Born Master Equation. As we mentioned in the main text, this corresponds to perform a non-self-consistent NCA calculation in which the propagator entering the self-energy is the bare emitter propagator rather than the fully dressed one. Therefore one expect that this approach is not able to capture the physics at strong system-bath couplings.  To see this we consider in Fig.~\ref{fig:born} the dynamics of the spontaneous emission  as in the main text and compare the NCA and Born Markov dynamics. In the left panel we consider the dynamics within RWA and for the emitter frequency on resonance with the lower edge of the photonic band. We show that the emergence of plateaux in the spontaneous emission, corresponding to freezing, are absent within Born-Markov. In fact this only features an exponential decay in time with a rate that grows with system-bath coupling. In the right panel we plot the dynamics at fixed value of the coupling $g$ and for different detuning $\Delta$. We note a large deviation for $\Delta=0$, an emitter frequency at the center of the band in correspondence of the van Hove singularity. Here the relxation dynamics is non-exponential and clearly cannot be captured by Born-Markov. Similar devations appear also for $\Delta=-W$, i.e. in correspondence of the lower edge of the photonic band. These result highlight the importance of the self-consistent dressing in our dynamical map, particularly at strong coupling.  We conclude by noting that a substantial difference between NCA and Born-Markov appears also in the spectral properties of the quantum emitter. In particular the large broadening of the resonance observed in Sec.~\ref{sec:results} cannot be reproduced with a Born-Markov calculation, whose spectral function always features a narrow in gap peak which does not broaden and only slightly shifts in frequency.

%
%
%
%
%

\end{document}